%% file: main.tex
\pdfoutput=1
\documentclass[12pt,a4paper]{article}

\usepackage{ifthen} 
\newboolean{pdflatex}
\setboolean{pdflatex}{true} 
\usepackage{ulem}
\newboolean{articletitles}
\setboolean{articletitles}{true} 

\newboolean{uprightparticles}
\setboolean{uprightparticles}{false} 

\def\paperauthors{LHCb collaboration} 
\def\paperasciititle{rhoPbPb} 
\def\papertitle{
Coherent photoproduction of $\rho^0, \omega$ and excited vector mesons in ultraperipheral PbPb
collisions
} 
\def\paperkeywords{{High Energy Physics}, {LHCb}} 
\def\papercopyright{\the\year\ CERN for the benefit of the LHCb collaboration} 
\def\paperlicence{CC BY 4.0 licence}
\def\paperlicenceurl{https://creativecommons.org/licenses/by/4.0/}

\newif\ifEnableSectionTOCLinks
\EnableSectionTOCLinksfalse 

\input{preamble}

\begin{document}

\renewcommand{\thefootnote}{\fnsymbol{footnote}}
\setcounter{footnote}{1}

\onecolumn
\input{title-LHCb-PAPER}


\renewcommand{\thefootnote}{\arabic{footnote}}
\setcounter{footnote}{0}

\cleardoublepage


\pagestyle{plain} 
\setcounter{page}{1}
\pagenumbering{arabic}


\input{body}
\input{acknowledgements}




\newpage
\addcontentsline{toc}{section}{References}
\bibliographystyle{LHCb}
\bibliography{main,standard,LHCb-PAPER,LHCb-CONF,LHCb-DP,LHCb-TDR}
 
\newpage
\input{Authorship_LHCb-PAPER-2024-042}

\end{document}

%% file: preamble.tex
\usepackage[top=1in, bottom=1.25in, left=1in, right=1in]{geometry}

\usepackage{microtype}
\usepackage{lineno}  
\usepackage{xspace} 
\usepackage{caption} 

\usepackage{graphicx}  
\usepackage{color}
\usepackage{colortbl}
\graphicspath{{./figs/}} 

\usepackage{amsmath} 
\usepackage{amssymb}
\usepackage{amsfonts}
\usepackage{upgreek} 

\newcommand*\patchAmsMathEnvironmentForLineno[1]{%
\expandafter\let\csname old#1\expandafter\endcsname\csname #1\endcsname
\expandafter\let\csname oldend#1\expandafter\endcsname\csname
end#1\endcsname
 \renewenvironment{#1}%
   {\linenomath\csname old#1\endcsname}%
   {\csname oldend#1\endcsname\endlinenomath}%
}
\newcommand*\patchBothAmsMathEnvironmentsForLineno[1]{%
  \patchAmsMathEnvironmentForLineno{#1}%
  \patchAmsMathEnvironmentForLineno{#1*}%
}
\AtBeginDocument{%
\patchBothAmsMathEnvironmentsForLineno{equation}%
\patchBothAmsMathEnvironmentsForLineno{align}%
\patchBothAmsMathEnvironmentsForLineno{flalign}%
\patchBothAmsMathEnvironmentsForLineno{alignat}%
\patchBothAmsMathEnvironmentsForLineno{gather}%
\patchBothAmsMathEnvironmentsForLineno{multline}%
\patchBothAmsMathEnvironmentsForLineno{eqnarray}%
}

\usepackage[pdftex,
            pdfauthor={\paperauthors},
            pdftitle={\paperasciititle},
            pdfkeywords={\paperkeywords}]{hyperref}
\usepackage{hyperxmp}
\hypersetup{
    pdfcopyright={Copyright (C) \papercopyright},
    pdflicenseurl={\paperlicenceurl}
}

\usepackage[colorinlistoftodos,textsize=scriptsize]{todonotes}

\usepackage[bottom,flushmargin,hang,multiple]{footmisc}

\usepackage[all]{hypcap} 

\input{lhcb-symbols-def} 

\hypersetup{
  colorlinks   = true, 
  urlcolor     = blue, 
  linkcolor    = blue, 
  citecolor    = red   
}

\ifEnableSectionTOCLinks
    \usepackage[explicit]{titlesec} 
    
    \let\oldcontentsline\contentsline
    \renewcommand

    \titleformat{\section}{\normalfont\Large\bf}{\hyperlink{tocsection.\thesection}{{\thesection} \parbox[t]{\dimexpr\textwidth-1pc}{#1}}}{1pc}{}

    \titleformat{\subsection}{\normalfont\bf}{\hyperlink{tocsubsection.\thesubsection}{{\thesubsection} \parbox[t]{\dimexpr\textwidth-1pc}{#1}}}{1pc}{}

    \titleformat{name=\section,numberless}[display]{}{}{0pt}{\normalfont\Huge\bfseries #1}
\fi

\usepackage{cite} 
\usepackage{mciteplus}

%% file: lhcb-symbols-def.tex
\usepackage{xspace} 
\usepackage{upgreek}

\def\lhcb   {\mbox{LHCb}\xspace}

\def\MagUp {\mbox{\em Mag\kern -0.05em Up}\xspace}

\ifthenelse{\boolean{uprightparticles}}%
{

 \def\Ppi         {\ensuremath{\uppi}\xspace}                 
                  
 \def\Prho        {\ensuremath{\uprho}\xspace}

 \def\Ppsi        {\ensuremath{\uppsi}\xspace}

 \def\PDelta      {\ensuremath{\Delta}\xspace}                 
 \def\PXi         {\ensuremath{\Xi}\xspace}                 
 \def\PLambda     {\ensuremath{\Lambda}\xspace}                 
 \def\PSigma      {\ensuremath{\Sigma}\xspace}                 
 \def\POmega      {\ensuremath{\Omega}\xspace}                 
 \def\PUpsilon    {\ensuremath{\Upsilon}\xspace}
 \let\oldPi\Pi
 \def\PPi         {\ensuremath{\oldPi}\xspace}

 \def\PB      {\ensuremath{\mathrm{B}}\xspace}                 
 \def\PD      {\ensuremath{\mathrm{D}}\xspace}                 
 \def\PJ      {\ensuremath{\mathrm{J}}\xspace}                 
 \def\PK      {\ensuremath{\mathrm{K}}\xspace}                 
 \def\Pb      {\ensuremath{\mathrm{b}}\xspace}                 
 \def\Pc      {\ensuremath{\mathrm{c}}\xspace}

 \def\Ps      {\ensuremath{\mathrm{s}}\xspace}

 \def\thebaroffset{0.0em}
}
{

 \def\Ppi         {\ensuremath{\pi}\xspace}                 
                  
 \def\Prho        {\ensuremath{\rho}\xspace}

 \def\Ppsi        {\ensuremath{\psi}\xspace}                 
                  
 \mathchardef\PDelta="7101
 \mathchardef\PXi="7104
 \mathchardef\PLambda="7103
 \mathchardef\PSigma="7106
 \mathchardef\POmega="710A
 \mathchardef\PUpsilon="7107
 \mathchardef\PPi="7105
 \def\PB      {\ensuremath{B}\xspace}                 
 \def\PD      {\ensuremath{D}\xspace}                 
 \def\PJ      {\ensuremath{J}\xspace}                 
 \def\PK      {\ensuremath{K}\xspace}                 
 \def\Pb      {\ensuremath{b}\xspace}                 
 \def\Pc      {\ensuremath{c}\xspace}

 \def\Ps      {\ensuremath{s}\xspace}

 \def\thebaroffset{0.18em}
}
\newcommand{\offsetoverline}[2][\thebaroffset]{\kern #1\overline{\kern -#1 #2}}%

\makeatletter
\ifcase \@ptsize \relax
  \newcommand{\miniscule}{\@setfontsize\miniscule{4}{5}}
\or
  \newcommand{\miniscule}{\@setfontsize\miniscule{5}{6}}
\or
  \newcommand{\miniscule}{\@setfontsize\miniscule{5}{6}}
\fi
\makeatother

\DeclareRobustCommand{\optbar}[1]{\shortstack{{\miniscule (\rule[.5ex]{1.25em}{.18mm})}
  \\ [-.7ex] $#1$}}

\def\squark    {{\ensuremath{\Ps}}\xspace}

\def\cquark    {{\ensuremath{\Pc}}\xspace}

\def\bquark    {{\ensuremath{\Pb}}\xspace}

\def\pion   {{\ensuremath{\Ppi}}\xspace}
\def\piz    {{\ensuremath{\pion^0}}\xspace}
\def\pip    {{\ensuremath{\pion^+}}\xspace}
\def\pim    {{\ensuremath{\pion^-}}\xspace}

\def\rhomeson {{\ensuremath{\Prho}}\xspace}
\def\rhoz     {{\ensuremath{\rhomeson^0}}\xspace}

\def\kaon    {{\ensuremath{\PK}}\xspace}

\def\KorKbar {\kern \thebaroffset\optbar{\kern -\thebaroffset \PK}{}\xspace}

\def\KS      {{\ensuremath{\kaon^0_{\mathrm{S}}}}\xspace}

\def\D       {{\ensuremath{\PD}}\xspace}

\def\DorDbar {\kern \thebaroffset\optbar{\kern -\thebaroffset \PD}\xspace}

\def\Dp      {{\ensuremath{\D^+}}\xspace}
\def\Dm      {{\ensuremath{\D^-}}\xspace}

\def\DpDm    {\ensuremath{\Dp {\kern -0.16em \Dm}}\xspace}

\def\B       {{\ensuremath{\PB}}\xspace}

\def\BorBbar {\kern \thebaroffset\optbar{\kern -\thebaroffset \PB}\xspace}

\def\Bd      {{\ensuremath{\B^0}}\xspace}

\def\BdorBdbar {\kern \thebaroffset\optbar{\kern -\thebaroffset \Bd}\xspace}

\def\Bs      {{\ensuremath{\B^0_\squark}}\xspace}

\def\BsorBsbar {\kern \thebaroffset\optbar{\kern -\thebaroffset \Bs}\xspace}

\def\jpsi     {{\ensuremath{{\PJ\mskip -3mu/\mskip -2mu\Ppsi}}}\xspace}

\def\Y#1S{\ensuremath{\PUpsilon{(#1S)}}\xspace}

\def\LorLbar     {\kern \thebaroffset\optbar{\kern -\thebaroffset \PLambda}\xspace}

\def\AT#1     {\ensuremath{A_{\mathrm{T}}^{#1}}\xspace}           

\def\C#1      {\ensuremath{\mathcal{C}_{#1}}\xspace}                       
\def\Cp#1     {\ensuremath{\mathcal{C}_{#1}^{'}}\xspace}                    
\def\Ceff#1   {\ensuremath{\mathcal{C}_{#1}^{\mathrm{(eff)}}}\xspace}        
\def\Cpeff#1  {\ensuremath{\mathcal{C}_{#1}^{'\mathrm{(eff)}}}\xspace}       
\def\Ope#1    {\ensuremath{\mathcal{O}_{#1}}\xspace}                       
\def\Opep#1   {\ensuremath{\mathcal{O}_{#1}^{'}}\xspace}                    

\newcommand{\nospaceunit}[1]{\ensuremath{\text{#1}}}       
\newcommand{\aunit}[1]{\ensuremath{\text{\,#1}}}       

\newcommand{\tev}{\aunit{Te\kern -0.1em V}\xspace}
\newcommand{\gev}{\aunit{Ge\kern -0.1em V}\xspace}
\newcommand{\mev}{\aunit{Me\kern -0.1em V}\xspace}
\newcommand{\kev}{\aunit{ke\kern -0.1em V}\xspace}
\newcommand{\ev}{\aunit{e\kern -0.1em V}\xspace}
 
\newcommand{\mevc}{\ensuremath{\aunit{Me\kern -0.1em V\!/}c}\xspace}
\newcommand{\gevc}{\ensuremath{\aunit{Ge\kern -0.1em V\!/}c}\xspace}
\newcommand{\mevcc}{\ensuremath{\aunit{Me\kern -0.1em V\!/}c^2}\xspace}
\newcommand{\gevcc}{\ensuremath{\aunit{Ge\kern -0.1em V\!/}c^2}\xspace}

\def\m    {\aunit{m}\xspace}

\def\fm   {\aunit{fm}\xspace}

\def\mub{\ensuremath{\,\upmu\nospaceunit{b}}\xspace}

\def\deriv {\ensuremath{\mathrm{d}}}

\def\gsim{{~\raise.15em\hbox{$>$}\kern-.85em
          \lower.35em\hbox{$\sim$}~}\xspace}
\def\lsim{{~\raise.15em\hbox{$<$}\kern-.85em
          \lower.35em\hbox{$\sim$}~}\xspace}

\def\sqsnn {\ensuremath{\protect\sqrt{s_{\scriptscriptstyle\text{NN}}}}\xspace}
\def\pt         {\ensuremath{p_{\mathrm{T}}}\xspace}

\def\tell1  {TELL1\xspace}
\def\ukl1   {UKL1\xspace}

\newcommand{\lhcborcid}[1]{\href{https://orcid.org/#1}{\hspace*{0.1em}\raisebox{-0.45ex}{\includegraphics[width=1em]{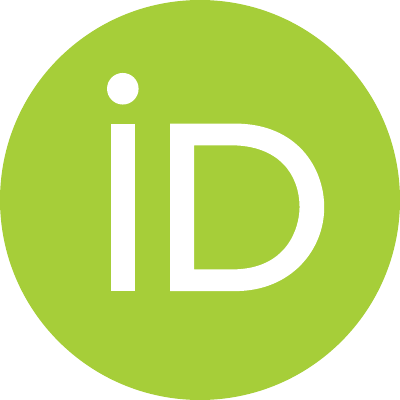}}}}


%% file: title-LHCb-PAPER.tex

\begin{titlepage}
\pagenumbering{roman}

\vspace*{-1.5cm}
\centerline{\large EUROPEAN ORGANIZATION FOR NUCLEAR RESEARCH (CERN)}
\vspace*{1.5cm}
\noindent
\begin{tabular*}{\linewidth}{lc@{\extracolsep{\fill}}r@{\extracolsep{0pt}}}
\vspace*{-1.5cm}\mbox{\!\!\!\includegraphics[width=.14\textwidth]{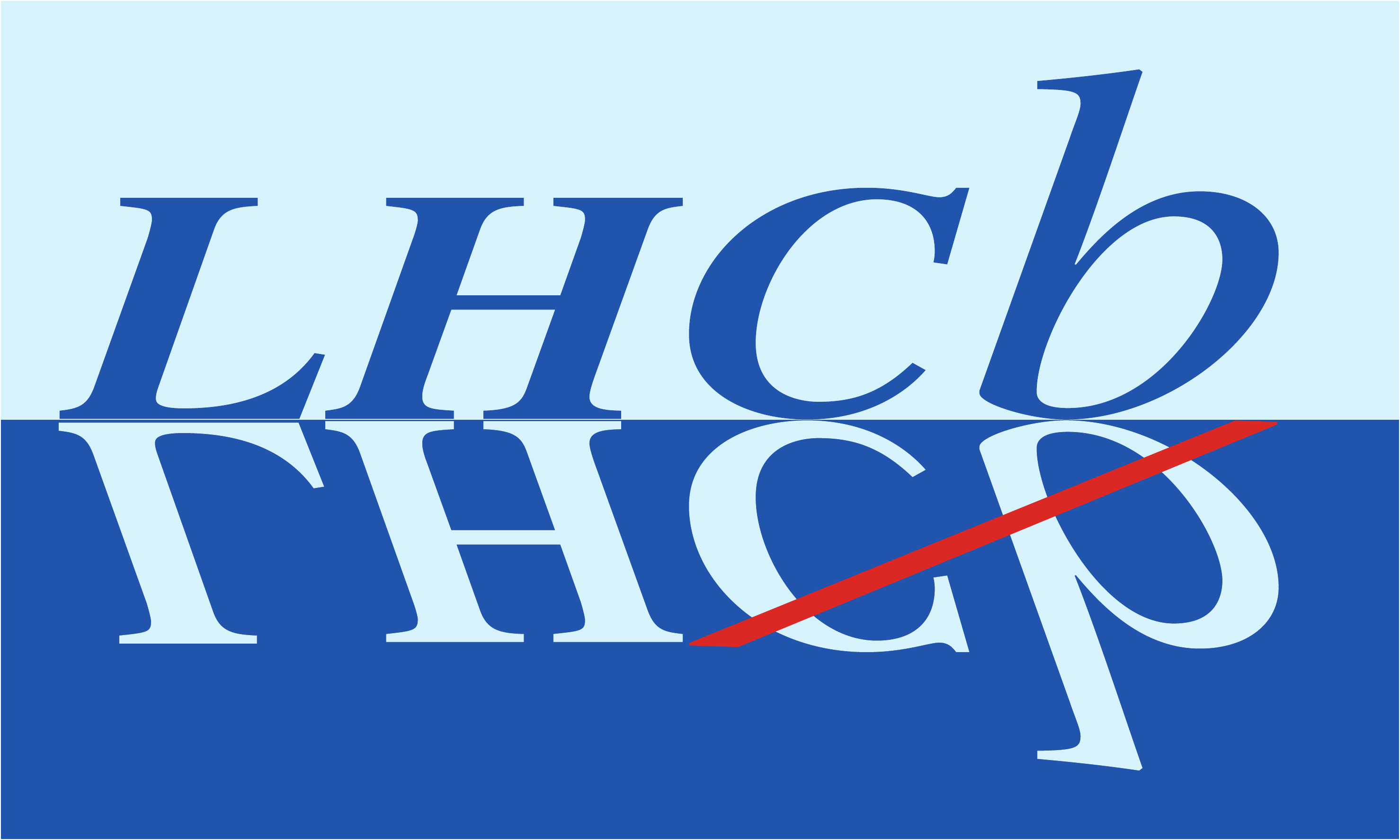}} & &%
\\
 & & CERN-EP-2025-106 \\  
 & & LHCb-PAPER-2024-042 \\  
 & & 
 18 November 2025 \\ 
 & & \\
\end{tabular*}

\vspace*{2.0cm}

{\normalfont\bfseries\boldmath\huge
\begin{center}
  \papertitle 
\end{center}
}

\vspace*{1.0cm}

\begin{center}
\paperauthors\footnote{Authors are listed at the end of this paper.}
\end{center}

\vspace{\fill}

\begin{abstract}
  \noindent
The invariant-mass distribution for the coherent photoproduction of dipions in ultraperipheral PbPb collisions is measured using
data, corresponding to an integrated luminosity of  $ 224.6 \pm 9.6 \mub^{-1}$, 
 collected by the \lhcb experiment in 2018 at a nucleon-nucleon centre-of-mass energy \mbox{$\sqsnn=5.02\tev$}.
In the mass range from 400 to 1200\mev, the results are consistent with previous experiments, with the spectrum dominated by the $\rhoz$ meson, which interferes with a nonresonant component, together with a smaller $\omega$ meson contribution.
In an extended mass range up to 2300\mev, models previously used do not fit the data and a consistent description requires the introduction of two resonances at masses of $1350\pm20\mev$ and $1790\pm20$\mev with widths of about 300\mev.
The cross-section for each meson is measured differentially in twelve bins of rapidity from 2.05 to 4.90. 
The $\rho^0$ cross-section increases with rapidity from about 400 to 600 mb and is measured with a typical precision of 8\%, while the cross-section times branching fraction for the $\omega,\rho^\prime$ and $\rho^{\prime\prime}$, with the statistical precision of the data, do not have a pronounced rapidity dependence and are between 0.5 and 1.5mb, with uncertainties up to 30\%.
A large nuclear suppression is observed for the $\rho^0$ meson compared to expectations based on photoproduction on the proton that use the impulse approximation. 
Significant suppression is also observed compared to that predicted by elastic scattering described in the Glauber approach, or with the addition of inelastic scattering in a Gribov--Glauber model.

\end{abstract}

\vspace*{1.cm}

\begin{center}
  Published in JHEP 11 (2025) 103.
\end{center}

\vspace{\fill}

{\footnotesize 
\centerline{\copyright~\papercopyright. \href{\paperlicenceurl}{\paperlicence}.}}
\vspace*{2mm}

\end{titlepage}


\newpage
\setcounter{page}{2}
\mbox{~}
%
%
%
%

%% file: body.tex
\section{Introduction}
\begin{figure}[b]
    \centering
        \includegraphics[scale=0.25]{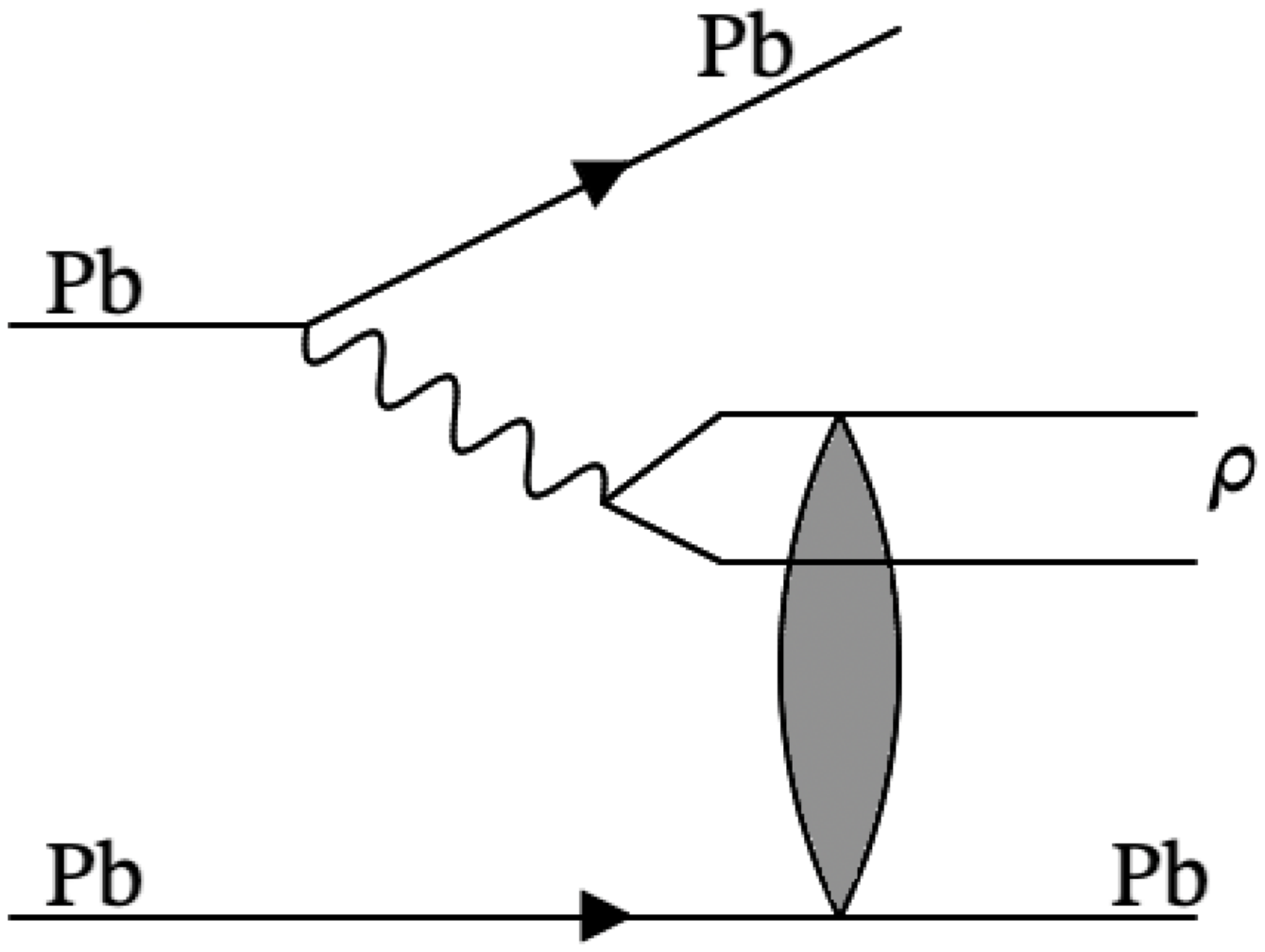}         
    \caption{Graphical representation of $\rho$-meson photoproduction in PbPb collisions.  The shaded area depicts the interaction of the $\rho$ meson with the target.}
    \label{fig:fd}
\end{figure}

Photoproduction of vector mesons on nuclei probes nuclear structure and is sensitive to nuclear shadowing and saturation effects~\cite{Brodsky:1994kf,  Klein:2019qfb, Guzey:2020ntc, Accardi:2012qut}.
The process can be studied in PbPb collisions at the LHC, when a photon is emitted from one lead ion, fluctuates into a virtual quark pair, and collides with the other target ion.
The interaction with the target is mediated by colourless QCD propagators, Reggeons or Pomerons~\cite{Donnachie:2002en}, and is shown diagrammatically in Fig.~\ref{fig:fd}.
Since the net colour flow in coherent photoproduction is zero, 
only a single vector meson is produced with a clear gap in rapidity relative to the  outgoing ions.
At small impact parameters,
additional strong interactions between the ions are likely, thereby destroying the rapidity gap. 
However, by moving beyond the range of the strong force, only electromagnetic sources need to be considered.
Experimentally, this is achieved in ultraperipheral collisions (UPC)~\cite{Bertulani:2005ru}, which produce the distinctive signature of a single reconstructed meson and no additional particle production, with the ions remaining intact.

This paper presents a study of the dipion final state produced in PbPb UPC, by the LHCb collaboration.
Although dominated by the $\rhoz$ meson, contributions are also present from the $\omega$ meson, nonresonant production, as well as possible excited vector mesons.
Measurements of the cross-sections for coherent $\rhoz$ and $\omega$ meson production are performed. 
The measured $\rhoz$ cross-section, $\sigma_{{\rm PbPb}\rightarrow {\rm Pb}\,\rhoz\, {\rm Pb}}$, together with a calculation of the photon flux, then allows the photoproduction cross-section on the nucleus, $\sigma_{\gamma {\rm Pb}\rightarrow \rhoz {\rm Pb}}$, to be derived.
A comparison of this result with photoproduction on the proton,
$\sigma_{\gamma p\rightarrow \rhoz p}$, allows the nuclear suppression factor to be determined~\cite{Guzey:2020ntc, Mantysaari:2023xcu}.
Na\"ively, in the impulse approximation~\cite{Chew:1952fca}, the ion-to-proton photoproduction ratio should scale as $A^{4/3}$ where $A$ is the number of nucleons.
However, since the $\rhoz$ meson is produced before it hits the target~\cite{Ioffe:1969kf}, it may undergo multiple interactions with the nucleons, which suppress the observable cross-section.  
This can be taken into account by the Gribov--Glauber mechanism~\cite{Gribov:1968jf,Glauber:1970jm} giving predictions~\cite{Guzey:2016piu} where the suppression depends on the total \mbox{$\rho$-nucleon} cross-section.  
The suppression for $\rhoz$ mesons is much larger than for $\jpsi$ mesons,  since the total $\rhoz$–nucleon cross-section is about 20\,mb, three order of magnitude greater than the \jpsi-nucleon cross-section, which leads to a suppression of $\sigma_{\gamma {\rm Pb}\rightarrow \rhoz {\rm Pb}}$ compared to the impulse approximation by a factor five for the \rhoz, in contrast to about 10\% for the \jpsi meson.
Suppression in excess of this would require additional mechanisms for nuclear shadowing~\cite{Guzey:2016qwo} such as saturation,
which is expected to be enhanced in nuclear collisions~\cite{Accardi:2012qut} and at low values of Bjorken-$x$, the fractional momentum of the partons in the nucleon.
In this respect, the forward region covered by the LHCb experiment is of particular interest since in $\rhoz$-meson photoproduction, $x$ values down to $10^{-6}$ are probed.

The $\rho$ meson is a broad resonance decaying almost exclusively to two pions.
Since its discovery in the early 1960s, there have been difficulties in measuring its mass and width.  
In photoproduction, the resonance is also skewed, which is attributed to sizeable interference with the nonresonant production of pion pairs.
A 1974 review by Spital and Yennie~\cite{Spital:1974cx} summarises the situation, which is essentially unchanged today, in observing the resonance shape from reconstructing $\rhoz\rightarrow\pip\pim$ decays.
 The most precise measurements of the mass and width come from $e^+e^-\rightarrow\pip\pim$ data and $\tau$ decays, where discrepancies between these measurements seem to have been resolved~\cite{Jegerlehner:2011ti}.
The quoted values in the PDG~\cite{PDG2024} have very small uncertainties on the mass, $m_\rhoz=775.26\pm0.23\mev$,\footnote{Natural units are used throughout this paper.} and width, $\Gamma_\rhoz=147.4\pm0.8\mev$, but there are several assumptions and model dependencies involved in this determination.  The situation is far from settled with a recent paper finding a value for the mass about 20\mev lower~\cite{Bartos:2017ils}.
Consistency in the $\rho$ parameters and the dipion spectrum measured in different production mechanisms is important, not just for an understanding of light-meson spectroscopy, but also because these feed into the determination of the hadronic corrections to the anomalous muon magnetic moment, $g-2$~\cite{Davier:2017zfy}.

The $\rho$ parameters can be determined very precisely in PbPb collisions with the LHCb detector, making use  of a large sample of about 20 million dipion events in UPC that has good invariant-mass resolution and is essentially free of background.
This is due to the excellent tracking reconstruction and particle identification of the LHCb detector and the fact that in UPC, the transverse momentum, \pt, being the Fourier transform of the impact parameter, is very small.

Although dominated by the $\rhoz$ meson and interference with nonresonantly-produced dipions, a precise measurement of the dipion spectrum allows other resonances to be identified.  
Interference between the $\rhoz$ and $\omega$ mesons, which was observed in AuAu collisions by the STAR collaboration~\cite{STAR:2017enh}, in $ep$ collisions by the H1 collaboration~\cite{H1:2020lzc}, and in $p$Pb collisions by the CMS collaboration~\cite{CMS:2019awk}, 
is seen allowing the
differential cross-section for $\omega$ production in UPC to be measured.
The dipion spectrum also indicates the presence of other particles, likely to be excited $\rhoz$ mesons.
In Ref.~\cite{ALICE:2020ugp}, the ALICE collaboration reports an excess of events in the high-mass region, corresponding to a resonance mass of $1725\pm17\mev$ and a width of $143\pm21\mev$, as determined with a simple Gaussian model. 
With a similar model, the STAR collaboration~\cite{Klein:2016dtn} obtained a mass of $1653\pm10\mev$ and a width of $164\pm15\mev$, which they noted was compatible with the spin-3 $\rho_3(1690)$.

Measurements of UPC $\rhoz$-meson production have been performed by the STAR collaboration~\cite{STAR:2007elq, STAR:2011wtm, STAR:2017enh}
at nucleon-nucleon centre-of-mass energies, $\sqrt{s_{\rm NN}}$, of 62.4 and 200\gev on gold nuclei, 
by the CMS collaboration in $p$Pb collisions at 5.02\tev, 
while the ALICE collaboration performed measurements on xenon at 5.44\tev~\cite{ALICE:2021jnv}, and lead at 2.76 and 5.02\tev~\cite{ALICE:2015nbw,ALICE:2020ugp,ALICE:2023jgu}.
The STAR and ALICE data on coherent $\rho$ production in heavy-ion collisions were compared to the impulse approximation , Glauber, and Gribov--Glauber models in Ref.\cite{Guzey:2016piu}, where it was shown that the  impulse approximation was about a factor four above the data, while the Glauber model was about a factor 1.5 too high.  With the implementation of a  Gribov--Glauber model, agreement with the data was achieved, although this used a parametrisation of H1 data that has since been improved~\cite{H1:2020lzc}.

Recent discussion on nuclear suppression in coherent production has revolved around measurement of the \jpsi meson, which have been performed by the STAR collaboration~\cite{STAR:2023vvb} and at the LHC by the ALICE~\cite{ALICE:2021tyx,ALICE:2021gpt,ALICE:2018oyo}, CMS~\cite{CMS:2023snh} and LHCb~\cite{LHCb-PAPER-2022-012} collaborations, where the nuclear suppression factor is consistent with, or exceeds, what is expected in saturation models~\cite{Mantysaari:2023xcu}.
Complementary measurements for the $\rho$ meson are welcome, since here the effect is much larger, although there are greater theoretical uncertainties when working in the nonperturbative regime. 

The structure of this paper is as follows. 
In Sec.~\ref{sec:data}, the detector, data-taking conditions and simulation  are described. The selection of the sample of dipions is detailed in Sec.~\ref{sec:sel}.
The lineshape measurement is presented in Sec.~\ref{sec:mass}.
The cross-section results are given in Sec.~\ref{sec:cs} and details of how these are used to extract the nuclear suppression factors are given in Sec.~\ref{sec:supp}.
Section~\ref{sec:conclude} presents conclusions.

%
%

\section{The LHCb detector, data collection and simulation}
\label{sec:data}

The LHCb detector~\cite{LHCb:2008vvz,LHCb:2014set} is a single-arm forward spectrometer covering the pseudorapidity range $2 < \eta < 5$, originally designed for the precision study of particles containing \bquark or \cquark quarks. 
The detector includes a high-precision tracking system consisting of a silicon-strip vertex detector surrounding the collision region, a large-area silicon-strip detector located upstream of a dipole magnet with a bending power of about 4\ T\,m, and three stations of silicon-strip detectors and straw drift tubes placed downstream of the magnet. 
The tracking system provides a measurement of the momentum, $p$, of charged particles with a relative uncertainty that varies from 0.5\% at low momentum to 1.0\% at 200\gev. 
The minimum distance of a track to a primary PbPb collision vertex is measured with a resolution of (15+29/\pt) $\mu$m, where \pt is in GeV. 
Different types of charged hadrons are distinguished using information from two ring-imaging Cherenkov (RICH) detectors. Photons, electrons and hadrons are identified by a calorimeter system consisting of scintillating-pad (SPD) and preshower detectors, an electromagnetic and a hadronic calorimeter. 
Muons are 
identified by a system composed of alternating layers of iron filters and
multiwire proportional chambers.
The pseudorapidity coverage is extended by forward and backward shower counters (HeRSCheL)
consisting of five planes of scintillators with three planes at 114.0, 19.7 and 7.5\m upstream
of the LHCb detector, and two planes downstream at 20.0 and 114.0\m. The HeRSCheL
detector~\cite{Akiba:2018neu}  significantly extends the acceptance for detecting particles from dissociated
nuclei by covering the pseudorapidity range of $5\lesssim|\eta|\lesssim10$, enhancing the classification
of central exclusive production and UPC events.
The online event selection is performed by a trigger, which consists of a hardware stage,
based on information from the calorimeter and muon systems, followed by a software
stage, which applies a full event reconstruction.

The data used in this analysis were collected in PbPb collisions in 2018 at a nucleon-nucleon centre-of-mass energy of 5.02\tev, which corresponds to a
total integrated luminosity of $224.6\pm9.6\mub^{-1}$~\cite{LHCb:2014vhh}.
The data were selected by a low-multiplicity trigger at the hardware stage that required between one and twenty SPD deposits, which has a signal efficiency above 98\% over most of the dipion mass and rapidity range, since only two SPD deposits are expected for the signal.
At the software trigger stage, one or more tracks were required to be reconstructed in detectors located both upstream and downstream of the magnet.

The use of simulation is limited to correcting for geometrical acceptance, measuring inefficiencies that amount to about $\sim3\%$, and constraining the yields of small backgrounds whose shapes are determined in data.
The production and decay of $\rhoz$ mesons is generated by the
\textsc{STARlight} program~\cite{Klein:2016yzr}, which is also used to investigate the two-photon production of muon and electron pairs.
In addition, since the kinematics of the decay are fully determined once the mass, rapidity and transverse momentum of the parent meson have been specified along with the decay angle in the rest-frame of the meson, the ROOT~\cite{Antcheva:2011zz} routine {\tt TGenPhaseSpace} is used to study other particle decays.
The interaction of the generated particles with the detector, and its response, are modelled using the Geant4 toolkit~\cite{GEANT4:2002zbu,Allison:2006ve} as described in Ref.~\cite{Clemencic:2011zza}.

%
%
\section{Selection of coherent UPC-produced dipions}
\label{sec:sel}

Coherent UPC, without the presence of additional hadronic interactions, can be selected by ensuring the impact parameter is greater than twice the radius of the ions and thus the transverse momentum of the central system is small, $\pt\sim 40\mev$, and the ions remaining intact, so no additional particles are produced apart from the central system.
For this analysis, the experimental signature is two pions and nothing else in the detector, with the pions being approximately back-to-back in the transverse plane.

Simulation is used to determine a fiducial region in which the detector reconstruction is efficient.  Both tracks must have $\pt>100\mev$ and $2<\eta<5$, and the parent meson must have a rapidity, $y$, in the range $2.05<y<4.90$
and a mass above 400\mev.
The angle, $\theta^*$, between the outgoing positively charged pion and the outgoing lead ion, in the rest frame of the parent meson, must satisfy $|\cos\theta^*|$ below a value that varies from 0.7, between rapidities of 3 and 4, to 0.2 at the edges of the detector.

Dipion candidates are formed by requiring precisely two oppositely charged tracks
reconstructed in the fiducial region with no additional tracks or photons.  The tracks must form a vertex that is spatially consistent with the average interaction region.
Both tracks must be consistent with the pion hypothesis created using information from the RICH detectors.
The contamination from dikaons in the sample is estimated to be below 0.5\% and is determined using $\phi\rightarrow K^+K^-$ decays that can be reconstructed in events where either track is consistent with the kaon hypothesis.
Neither track should be identified as a muon or electron.
For rapidities above 3.25, this reduces the contamination from the $\gamma\gamma\rightarrow ee$ and $\gamma\gamma\rightarrow\mu\mu$ processes to below 0.5\%, while
below rapidities of 3.25, the maximum contamination is 8\% and concentrated at the lowest dipion masses.
Low activity is required in the HeRSCheL detector, which has a good efficiency to detect nuclear breakup.
To suppress multihadron contamination including $\omega\rightarrow\pip\pim\piz$ and $\rhoz\rightarrow\pip\pim\gamma$ decays, it is required that the transverse momentum of the dipion system is below 100\mev, which reduces this background to below 0.5\%.
The final sample contains about 20 million candidates.

Acceptance and efficiency correction factors, $\varepsilon$, are determined as a function of the dipion mass, rapidity and $\cos\theta^*$, assuming that the $\rho$ meson inherits the transverse polarisation of the photon and decays following a $1-\cos^2\theta^*$ distribution.
The acceptance is mainly a geometric effect and is determined in simulation as the fraction of mesons decaying inside the fiducial volume, taking account of the magnetic field that can bend low-momentum particles out of the active region of the detector.  
At central rapidities it is close to 70\% decreasing to 5\% in the far-forward region.
A systematic uncertainty of between 4 and 6\%, which is the largest single source of uncertainty, is assigned  to this estimation by comparing the response of known pions from the decay $\KS\rightarrow\pip\pim$ in data and simulation.

The  $\KS\rightarrow\pip\pim$ sample in data is also used to determine the pion-identification efficiency, which is typically $(90\pm3)\%$ where the uncertainty arises from kinematic differences between the $\rho$ and $\KS$ mesons.
The efficiency due to the requirement that there are no photons reconstructed in the event is calculated per rapidity bin, using events with identified photons but a well-balanced \pt between the two pions.   This is typically 94\% and a systematic uncertainty of 2\% is assigned as the largest difference with simulation.
The vetoes on identified electrons or muons have typical efficiencies of $(98\pm1)\%$.
The efficiency for the HeRSCheL veto on activity varies from 80\% to 85\% with rapidity.  
That veto is estimated to remove about half of the events having an additional electromagnetic interaction between the nuclei~\cite{Guzey:2013jaa}, which leads to excitation of one or both ions, a subsequent neutron emission (seen using zero-degree calorimeters by the ALICE and CMS~\cite{CMS:2023snh,ALICE:2023jgu} collaborations), and partial or full breakup of the nucleus.
Such events are identified by the presence of a diffractive peak in the \pt distribution of the vetoed events and a systematic uncertainty of 2\% is assigned, reflecting the precision with which the shape of the incoherent component is determined. 
Efficiencies for the other selection requirements are above 94\%.
The total systematic uncertainty on the efficiency of the selection is about 7\%  of which 4\% is uncorrelated between rapidity bins.

%
%

\section{Fit to the dipion spectrum}
\label{sec:mass}

The fractions of dipions associated to the decay of the $\rho$ meson and to other processes are found from a $\chi^2$ fit to the  number of events, $N$, binned in dipion invariant mass, $m_{\pi\pi}$.
The data are fitted with a function of the form
\begin{equation}
\int \varepsilon(m'){\cal S}(m',\alpha_i)R(m_{\pi\pi},m')\deriv m' + {\cal B}(m_{\pi\pi}),
\end{equation}
where $\cal S$ is the signal function with free parameters $\alpha_i$, $\varepsilon$ is the efficiency and acceptance correction, and $R$ is a resolution function obtained from simulation and parametrised by the sum of two Gaussian functions.
The narrower Gaussian is used to describe dipion candidates whose associated tracks are fully reconstructed by the tracking system and has a width that varies linearly from 3.5 to 8.0\mev with rapidity and contributes a fraction that varies from 50\% to 95\% with rapidity.  The wider Gaussian has a width of 50\mev and accounts for candidates where one of the pions does not reach the tracking stations downstream of the magnet, usually due to low-momentum particles being swept out of the acceptance by the magnetic field.
The background contributes less than 1\% for $y>3.25$ and its shape, $\cal B$, is described using simulation and control samples in which 
the respective individual background source has been enhanced.
For $y<3.25$ the background is only significant at the lowest dipion masses: above 600\mev it is less than 3\%.

For the signal function, several different parametrisations are used.  First, a model, due to S\"oding~\cite{Soding:1965nh} that has contributions for a $\rho$ resonance and a nonresonant continuum that is constant with mass.  Second, an extension to this that includes a contribution for the $\omega$ resonance.  The third model uses a more flexible shape to describe the continuum.  Finally, two additional resonances are introduced to describe the spectrum in an extended mass range.

The first parametrisation of the dipion spectrum is due to S\"oding~\cite{Soding:1965nh}, 
which models the $\rho$ meson using a Breit--Wigner function with mass, $m_\rho$, and an energy-dependent width, $\Gamma$, and includes interference with a nonresonant contribution described using a constant:
\begin{equation}
{\cal S}(m_{\pi\pi}) 
\propto
\left|
f_\rho 
\frac{\sqrt{m_{\pi\pi}m_\rho\Gamma(m_{\pi\pi})}}
{m_{\pi\pi}^2-m_\rho^2+im_\rho\Gamma(m_{\pi\pi})}
+f_{\rm nr}
\right|
^2, 
~\label{eq:sod1}
\end{equation}
with 
\begin{equation}
  \Gamma(m_{\pi\pi})=\Gamma_\rho \frac{m_\rho}{m_{\pi\pi}}
    \frac{q(m_{\pi\pi})}{q(m_{\rho})},
    \label{eq:edepwidth}
\end{equation}
where $\Gamma_\rho$ is the pole width, $q(m)=\sqrt{m^2-4m_\pi^2}$ is the
momentum of the pion with mass $m_\pi$ in the centre-of-mass of a system with mass $m$,
while $f_\rho$ and $f_{\rm nr}$ are free parameters, describing the $\rho$ and nonresonant contributions, respectively.

\begin{table}[tb]
\caption{\small Results of the fit to LHCb data that uses Eq.~\ref{eq:sod1}, compared to measurements by other experiments.  
The uncertainties on the LHCb values are dominated by systematic effects.}
\resizebox{\columnwidth}{!}{
\begin{tabular}{l|c|c|c|c}

  Experiment&
  LHCb &
  ZEUS~\cite{ZEUS:1997rof}&
  H1~\cite{H1:2009cml,H1:1996prv}
  &
  ALICE~\cite{ALICE:2015nbw}
   \\
  \hline
  Fit range [\gev]&
  [0.4,1.2]&
  [0.55,1.20]&
  [0.6,1.1] &
  [0.6,1.5]
  
  \\

  $m_\rho$ [\mev]  &
$771\pm3$&
$770\pm2$&
$769\pm4$&
  $762.0^{+6.5}_{-3.8}$\\

$\Gamma_\rho$ [\mev]& 
$149\pm4$ &
$146\pm13$&
$162\pm8$& 
$150^{+13}_{-7}$

\\

$|f_{\rm nr}/f_\rho|$ [$\gev^{-{1}/{2}}]$&
$0.72\pm0.04$&
$0.70\pm0.04$&
$0.57\pm0.09$&
$0.50^{+0.10}_{-0.06}$ 
\\
\end{tabular}
}
\label{tab:simple}
 \end{table}

The fit is performed for $m_{\pi\pi}$ between 400\mev and 1200\mev in twelve bins of rapidity.  
Compared to measurements at other experiments~\cite{ZEUS:1997rof,H1:1996prv,H1:2009cml,H1:2020lzc,STAR:2007elq,STAR:2011wtm,STAR:2017enh, CMS:2019awk,ALICE:2015nbw, ALICE:2020ugp}, it is possible to extend the fit to lower mass values due to a reduced background contamination and the ability to reconstruct tracks down to a \pt of 100\mev.  
The results of the fit for the rapidity range $2.75<y<4.50$ are given in Table~\ref{tab:simple} and are similar to those from other experiments for the mass and width, while the nonresonant contribution differs between experiments, most likely due to different amounts of residual background.
Due to the large signal sample, the statistical uncertainties on the fit parameters are below 0.1\%.  
A systematic uncertainty is estimated from the standard deviation of the results in the individual rapidity bins and by fixing the background contribution to zero.

\begin{figure}[tb]
    \centering      
        \includegraphics[scale=0.5]{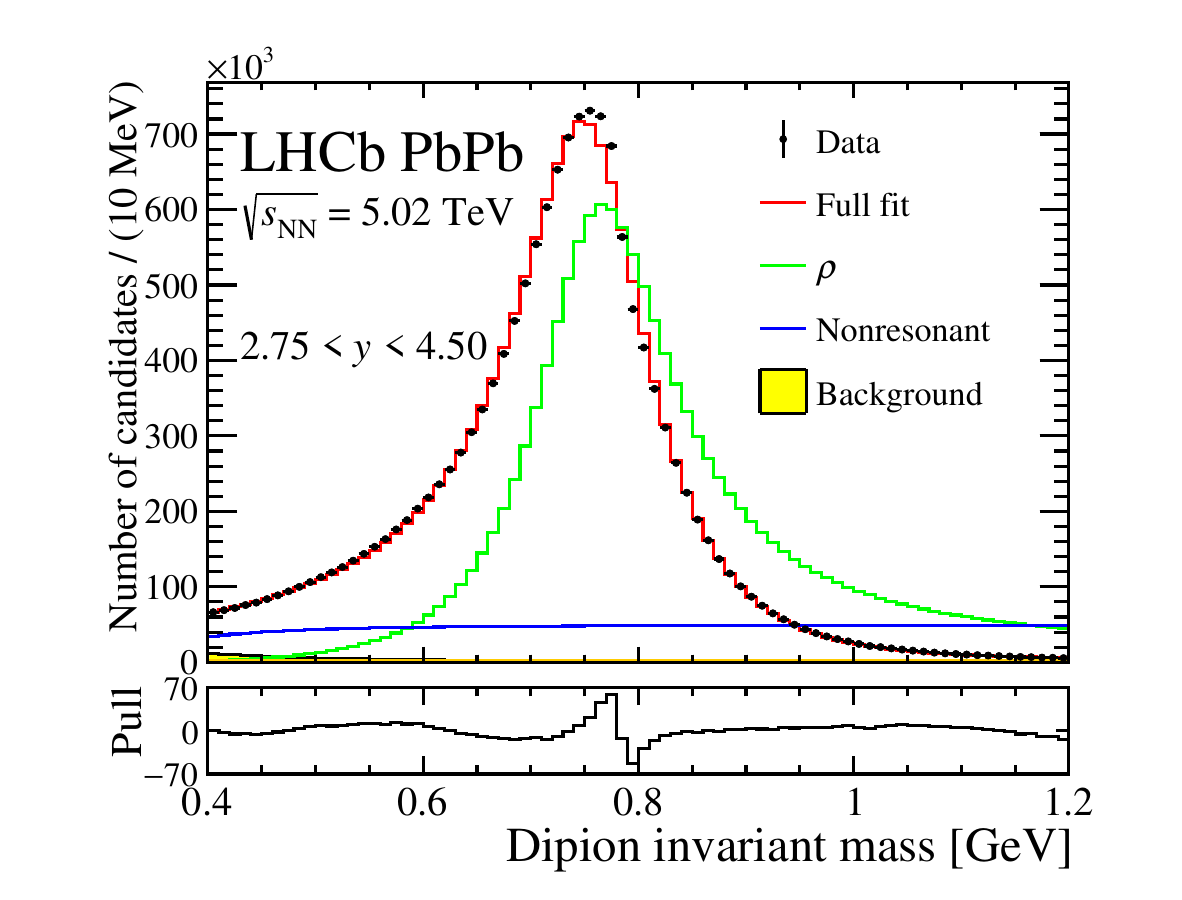}         
    \caption{Dipion invariant-mass distribution with the result of the fit superimposed, where the signal is modelled by the function in Eq.~\ref{eq:sod1}.
    The lower panel shows the normalised residuals.
}
    \label{fig:mass}
\end{figure}

\begin{figure}[tb]
\advance\leftskip-1cm
\advance\rightskip-2cm
    \includegraphics[scale=0.45]{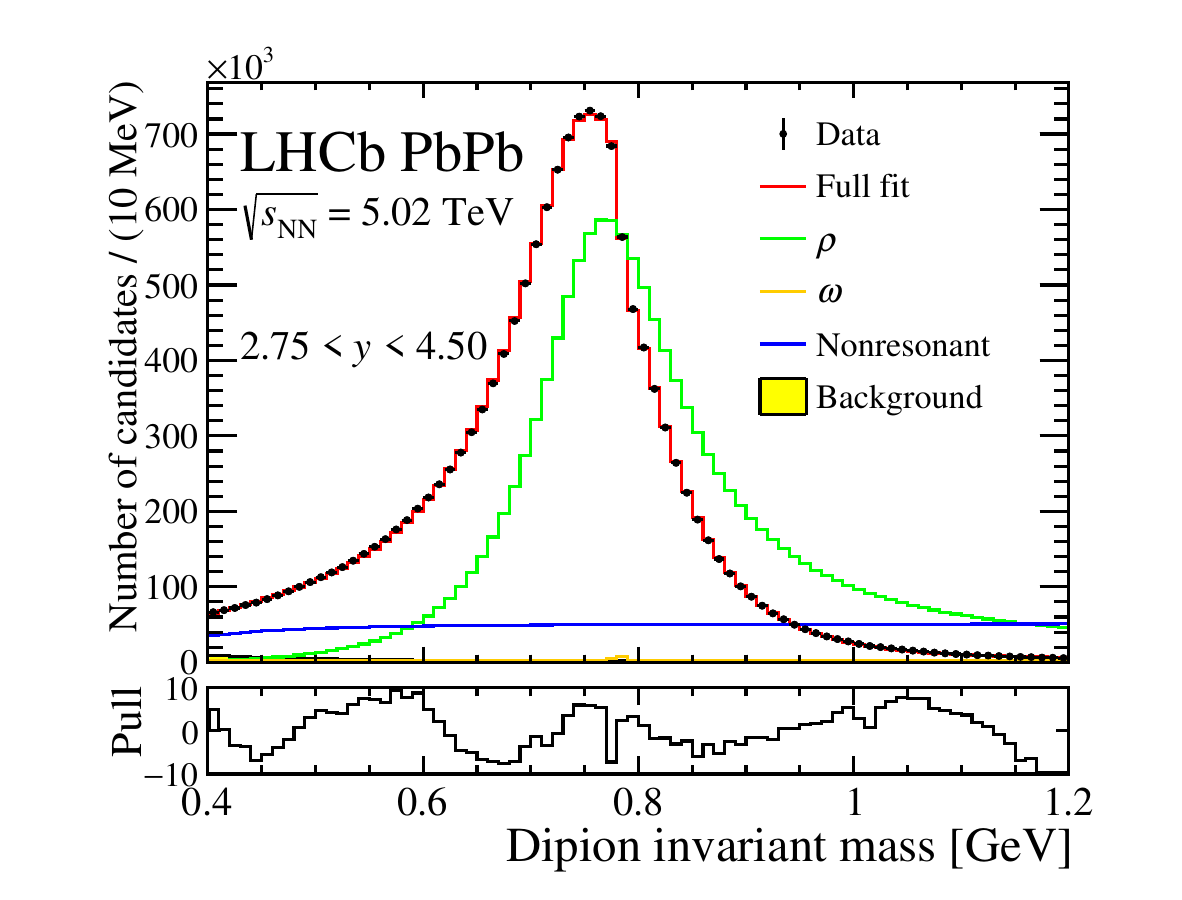}    
    \hspace*{-0.75cm}
    \includegraphics[scale=0.45]{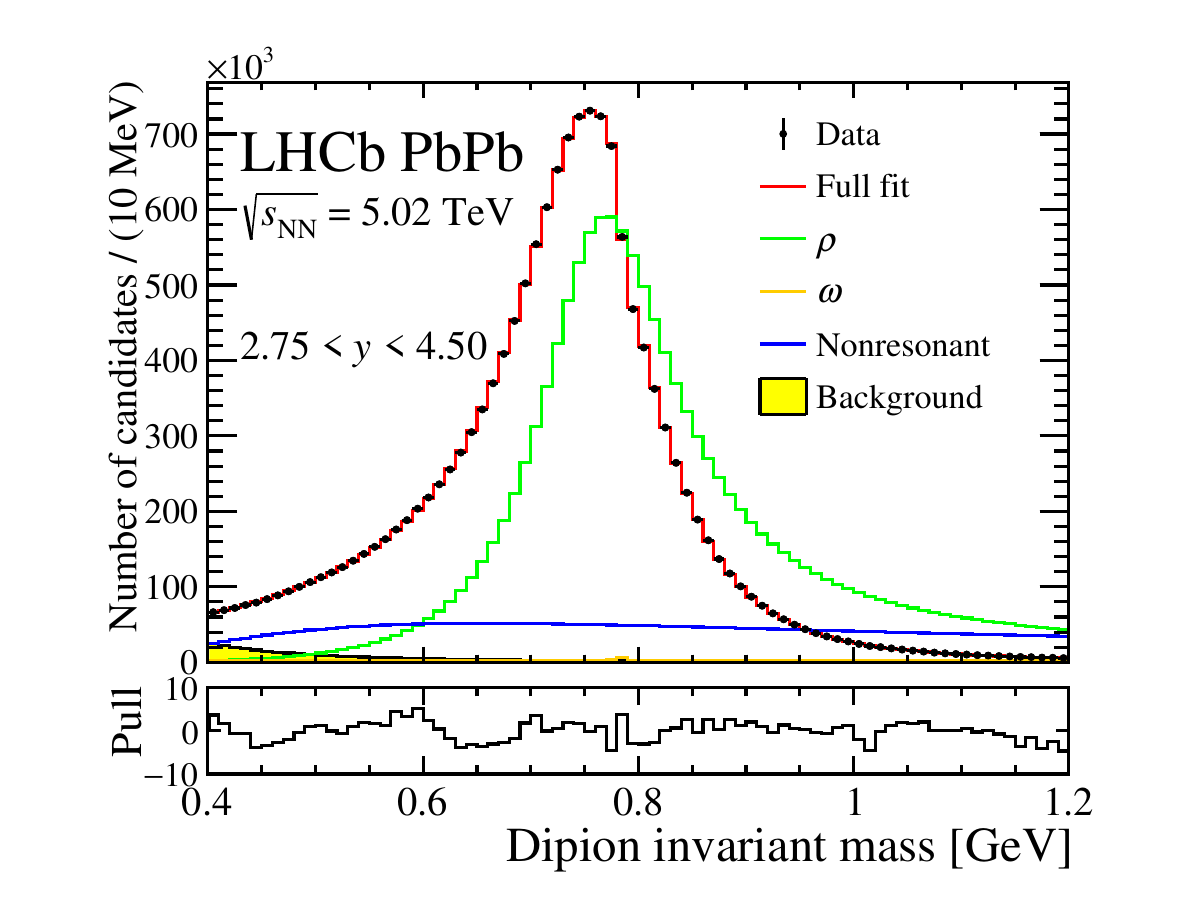}         
    \caption{Invariant-mass fits to data where the signal model includes a contribution for the $\omega$ resonance with the parametrisation of (left) Eq.~\ref{eq:sod2} and (right) Eq.~\ref{eq:sodh1}.
}
    \label{fig:withomega}
\end{figure}

The fit is shown in Fig.~\ref{fig:mass}.
Although the overall shape is similar to the data, the fit
quality is poor, particularly in the region of the $\omega$ resonance.
Similar observations were made by the STAR~\cite{STAR:2011wtm}, H1~\cite{H1:2020lzc}, and CMS~\cite{CMS:2019awk} collaborations and led to the introduction of a second Breit--Wigner function.  Following the STAR parametrisation, the signal function is extended to read
\begin{equation}
{\cal S}(m_{\pi\pi})= 
\left|
f_\rho\frac{\sqrt{m_{\pi\pi}m_\rho\Gamma(m_{\pi\pi})}}
{m_{\pi\pi}^2-m_\rho^2+im_\rho\Gamma(m_{\pi\pi})}
+f_{\rm nr}
+f_\omega \exp(i\phi_\omega)
\frac{\sqrt{m_{\pi\pi}m_\omega\Gamma_\omega}}
{
m_{\pi\pi}^2-m_\omega^2+im_\omega\Gamma_\omega}
\right|^2,
~\label{eq:sod2}
\end{equation}
where $f_\omega$ is the contribution of the $\omega$ resonance and $\phi_\omega$ is its phase relative to the $\rho$ meson.
The mass, $m_\omega$, and width, $\Gamma_\omega$, are fixed to the PDG values~\cite{PDG2024} and  
a constant width is assumed for $\Gamma_\omega$ as it is very narrow: implementing an energy-dependent width for the $\omega$ resonance makes minimal difference to the fit results.
The left panel of Fig.~\ref{fig:withomega} shows that the quality of the fit improves considerably in the peak region, and although the absolute size of the $\omega$ contribution is small, its presence and the resulting $\rho-\omega$ interference has a significant impact.
The parameters of the fit are presented in Table~\ref{tab:withomega1} and are consistent with the STAR results.

\begin{table}[tb]
  \caption{\small Results of fits to LHCb data that include an $\omega$ resonance using the parametrisation of Eq.~\ref{eq:sod2}, compared to measurements by other experiments.
  The uncertainties quoted for the LHCb measurements are the estimated systematic contributions.
}
\centering
  \begin{tabular}{l|r@{\:$\pm$\:}l|r@{\:$\pm$\:}l|r@{\:$\pm$\:}l}
    
    &\multicolumn{2}{c|}{LHCb}&\multicolumn{2}{c|}{STAR~\cite{STAR:2017enh}}
    &
    \multicolumn{2}{c}{
      CMS~\cite{CMS:2019awk}}
    \\

\hline
$m_\rho$ [\mev] & 774&3 &  
776.2&0.2
&773&1
\\
$\Gamma_\rho$ [\mev]& 153&3 &  
148&1 &
148&3
\\
$|f_{\rm nr}/f_\rho| [\gev^{-1/2}]$ & 0.73&0.03 &
0.79&0.08  
&0.50&0.06  
\\
$\phi_\omega$ [rad] & 1.4&0.1 &  
1.46&0.11
&1.8&0.3
 \\
$|f_\omega/f_\rho| $ & 0.32&0.03 &    0.36&0.05  &
0.40&0.06
\\

\end{tabular}

\label{tab:withomega1}
\end{table}

\begin{table}[tb]
  \caption{\small Results of fits to LHCb data that include an $\omega$ resonance using the parametrisation of Eq.~\ref{eq:sodh1}, compared to the measurement by the H1 collaboration.
  The uncertainties quoted for the LHCb measurements are the estimated systematic contributions.
}
\centering
  \begin{tabular}{l|r@{\:$\pm$\:}l|r@{\:$\pm$\:}l}
    &\multicolumn{2}{c|}{LHCb}&
    \multicolumn{2}{c}{H1~\cite{H1:2020lzc}}\\
    
\hline
$m_\rho$ [\mev] &  774&3& 
770.8&$1.3^{+2.3}_{-2.4}$ 
\\
$\Gamma_\rho$ [\mev]&  153&3 & 
151.3&$2.2^{+1.6}_{-2.8}$
\\
$|f_{\rm nr}/f_\rho| $ &  0.18&0.02 
&0.189&$0.026^{+0.025}_{-0.016}$
\\
$\phi_\omega$ [rad] &  $-0.20$&0.04 
& $-0.53$&$0.22^{+0.21}_{-0.17}$ 
 \\
$|f_\omega/f_\rho| $ & 0.15&0.01&  
0.166&$0.017^{+0.008}_{-0.023}$  \\
$\Lambda$ [\mev] & 
  
400&100& 
180&$590^{+200}_{-100}$
\\
$\delta$&1.1&0.1  
& 0.76&$0.35^{+0.14}_{-0.08}$
\\
\end{tabular}

\label{tab:withomega}
\end{table}

The H1 collaboration used a different parametrisation of the phase and interference
and also included two additional parameters to allow the nonresonant  shape to differ from a constant: 
\begin{equation}
\textstyle
{\cal S}(m_{\pi\pi})=
\frac{q^3(m_{\pi\pi})}{q^3(m_\rho)}
\left|
{f_\rho\,\cal{BW}}_\rho(m_{\pi\pi})
\biggl(
1+f_\omega\exp{({i\phi_\omega})}
\frac{m_{\pi\pi}^2}{ m_\omega^2}
{\cal{BW}}_\omega(m_{\pi\pi})
\biggr)
+\frac{f_{\rm nr}}
{(m_{\pi\pi}^2-4m_\pi^2+\Lambda^2)^\delta}
\right|
^2\, ,
~\label{eq:sodh1}
\end{equation}
where 
\begin{equation}
{\cal{BW}}_V(m_{\pi\pi})=\frac{m_V\Gamma_V}{m_{\pi\pi}^2-m_V^2+im_V\Gamma(m_{\pi\pi})} \, ,
\end{equation}
are Breit--Wigner distributions for the $\rho$ and
$\omega$ resonances (with $V=\rho,\omega)$, and $\Lambda$ and $\delta$ are free parameters.
Adopting their convention gives a better fit to the data as can be seen in the right panel of Fig.~\ref{fig:withomega}.
The results are summarised in Table~\ref{tab:withomega} and are consistent with the H1 results.

\begin{figure}[tb]
\advance\leftskip-1cm
\advance\rightskip-2cm
        \includegraphics[scale=0.45]{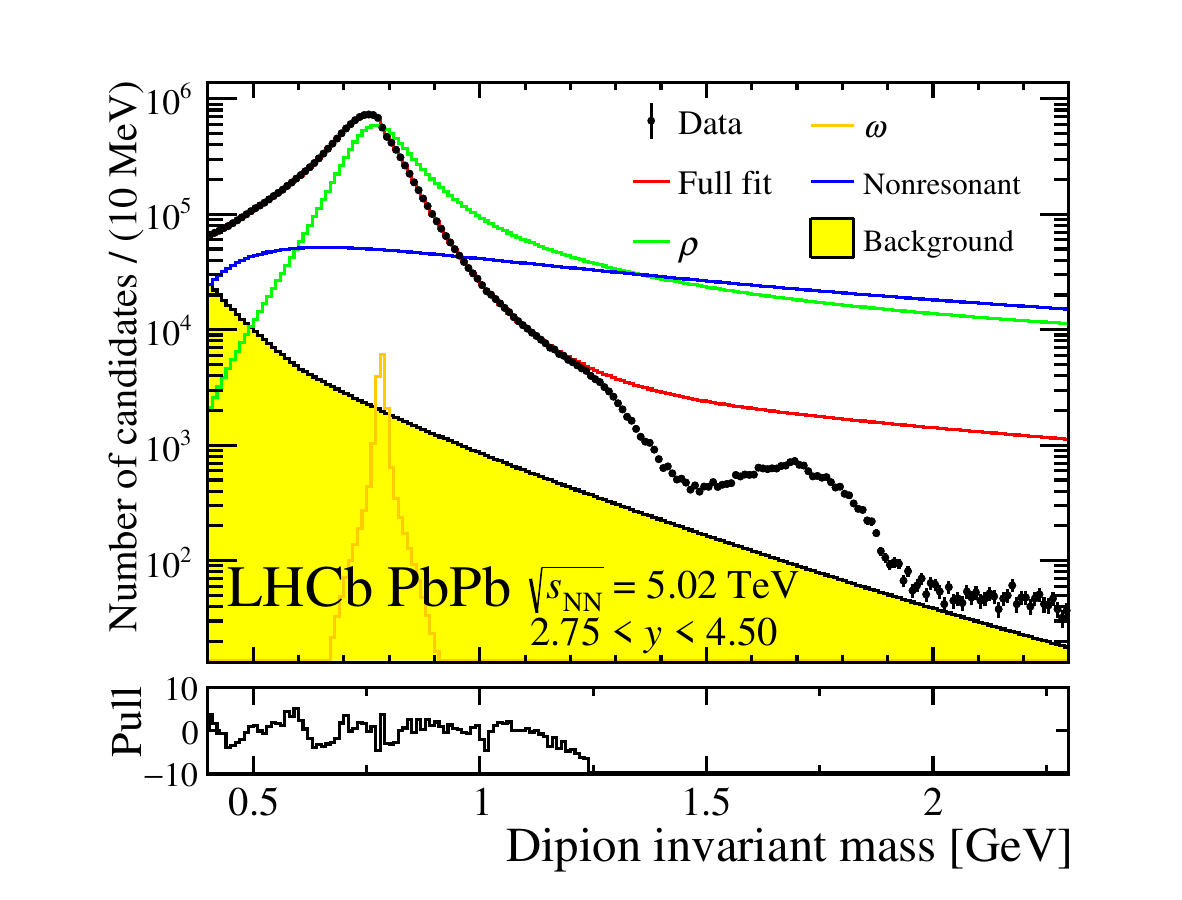} 
            \hspace*{-0.75cm}
        \includegraphics[scale=0.45]{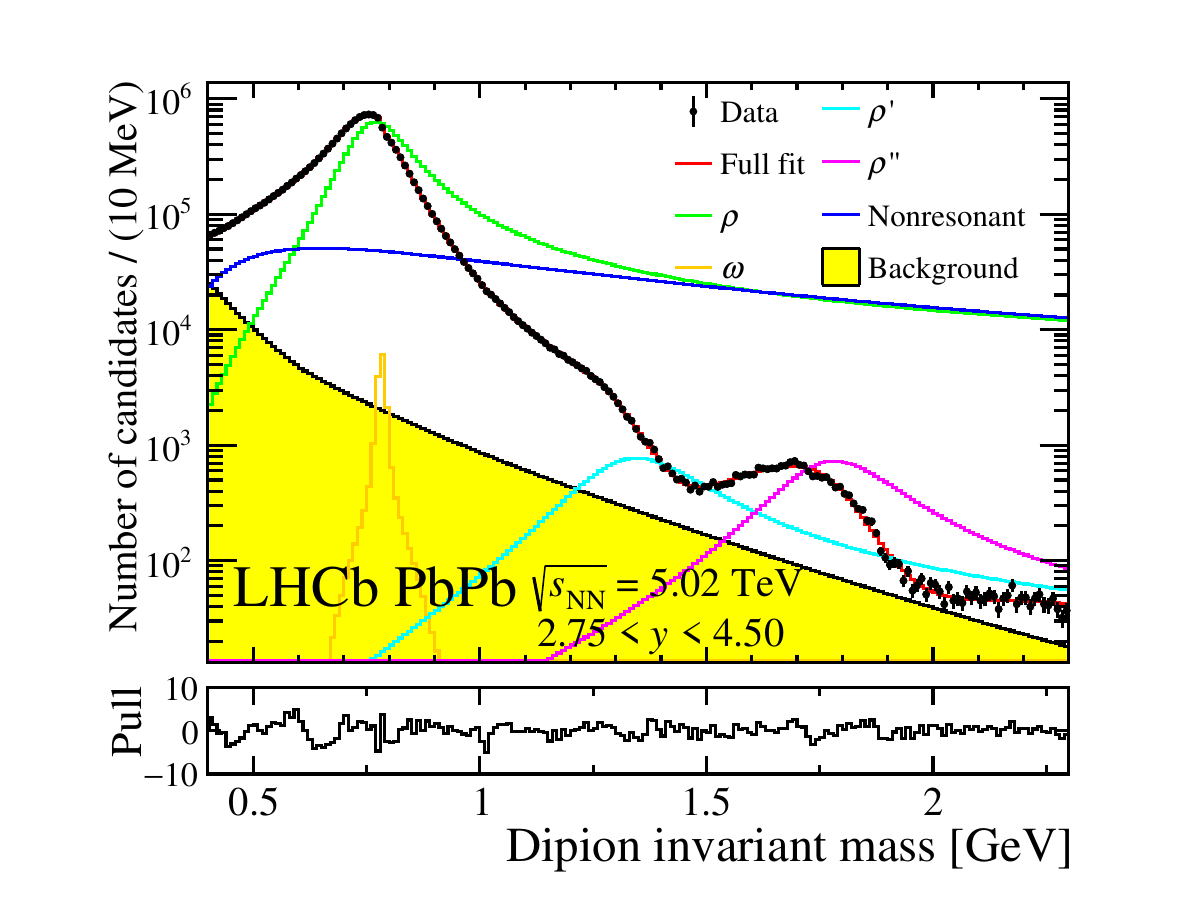}
    \caption{
    Invariant-mass fits to data (left) 
    in the region between 400 and 1200\mev,
    where the signal contribution is parametrised by Eq.~\ref{eq:sodh1},
    and (right) in the full region, where the signal model is parametrised by Eq.~\ref{eq:best}, which includes two additional resonances and a phase for the nonresonant component.
    }
    \label{fig:rhomass_full_log}
\end{figure}

Although the fit is acceptable in the fitted region, when expanding the fit range up to dipion masses of 2.3\gev, displayed in the left panel of Fig.~\ref{fig:rhomass_full_log},
the model overshoots the data by an order of magnitude.  
Furthermore, the data show a clear structure at around 1.7\gev, which is also  discernible in STAR~\cite{Klein:2016dtn} and ALICE~\cite{ALICE:2020ugp} data.
To improve the quality of the fit to the extended mass range,
two modifications are made to the signal model that now takes the form:
\begin{equation}
\begin{split}
{\cal S}(m_{\pi\pi})=
&
\frac{q^3(m_{\pi\pi})}{q^3(m_\rho)}
\Bigg|
f_\rho\, 
{\cal{BW}}_\rho(m_{\pi\pi})
\biggl(
1+f_\omega\exp({i\phi_\omega})
\frac{m_{\pi\pi}^2}{m_\omega^2}
{\cal{BW}}_\omega(m_{\pi\pi})
\biggr)\\
&+\frac{f_{\rm nr}\exp(i\phi_{\rm nr})}
{(m_{\pi\pi}^2-4m_\pi^2+\Lambda^2)^\delta}
+f_{\rho\prime} \exp({i\phi_{\rho\prime}})
{\cal{BW}}_{\rho\prime}(m_{\pi\pi})
+f_{\rho\prime\prime} \exp({i\phi_{\rho\prime\prime}})
{\cal{BW}}_{\rho\prime\prime}(m_{\pi\pi})
\Bigg|^2\, .\\
\end{split}
~\label{eq:best}
\end{equation}
First, a phase factor, $\phi_{\rm nr}$, is introduced between the $\rho$ and nonresonant components and this leads to a reduction of the signal function at higher masses.
Second, to describe the bump around 1.7\gev and the cusp at about 1.2\gev, two extra Breit--Wigner resonances are introduced to describe excited $\rho$ states, designated as $\rho^\prime$ and $\rho^{\prime\prime}$, with $f_{\rho\prime}$ and
$f_{\rho\prime\prime}$ the contributions of each.
The parameters of the fit are given in Table~\ref{tab:rho_r12} and the 
result is shown in the right panel of Fig.~\ref{fig:rhomass_full_log}.
Excellent agreement with data is obtained across the full range.

There is a long history of possible vector states in the high-mass region as discussed in the PDG review \cite{PDG2024}.  
The $\rho^\prime$ resonance is measured to have a mass of $1350\pm5\pm20\mev$ and a width of $320\pm10\pm40\mev$, while the 
$\rho^{\prime\prime}$ resonance has a mass of
\mbox{$1790\pm5\pm20\mev$} and a width of $290\pm10\pm40\mev$, where the first uncertainty is the statistical precision of the fit result and the second is due to systematic effects, evaluated using the standard deviation of results obtained by fitting the model separately in each rapidity bin.
The statistical significance of either or both resonances compared to the null hypothesis of Eq.~\ref{eq:sodh1} greatly exceeds the threshold for claiming an observation.
However, it should be pointed out that improvements to the fit have been obtained with alternative parametrisations including using a form factor to multiply $\cal{BW}_\rho$, and including a contribution from a noninterfering $f_2(1270)$ meson.  

These measurements are consistent with the values of the $\rho(1450)$ and $\rho(1700)$ mesons in the PDG~\cite{PDG2024}, noting the role the choice of dynamic lineshapes can play.
Further evidence for the existence of these particles was reported by the Babar collaboration using $e^+e^-\rightarrow\pi\pi(\gamma)$ data~\cite{BaBar:2012bdw} where
the same qualitative features are observed as in the LHCb data.
The production processes are similar with
the main difference being that in photoproduction the photon is quasi-real while it is strongly virtual in $e^+e^-$ annihilation.  
The coupling between the photon and the pions is probed in both processes so the LHCb data may help clarify the resonant contributions, which have an impact in determining  the hadronic corrections to the $g-2$ measurement~\cite{Davier:2019can}.

\begin{table}[tb]
  \caption{\small Result of the fit to data with the signal model of 
  Eq.~\ref{eq:best}, which includes a phase for the nonresonant component, and Breit--Wigner functions for the $\omega$, $\rho^\prime$ and $\rho^{\prime\prime}$ resonances.
  The uncertainties are dominated by systematic effects.}
\centering
  \begin{tabular}{l|r@{\:$\pm$\:}l}
    
    Parameter (Eq.~\ref{eq:best}) & \multicolumn{2}{c}{Result}  \\
\hline
$m_\rho$ [\mev] & 774&3 \\
$\Gamma_\rho$ [\mev]& 148&7 \\
$|f_{\rm nr}/f_\rho|$ & 0.18&0.03\\
$\phi_{\rm nr}$ & $-0.1$&0.1\\
$\Lambda$ [\mev] & 420&150\\
$\delta$ & 1.2&0.1 \\
$|f_\omega/f_\rho|$ & 0.14&0.02\\
$\phi_\omega$ [\,rad\,] & $-0.20$&0.06\\
$|f_{\rho\prime}/f_\rho|$  &  0.013&0.008\\
$m_{\rho\prime}$ [\mev] & 1350&20 \\
$\Gamma_{\rho\prime}$ [\mev] &320&40 \\     
$\phi_{\rho^\prime}$ & 2.4&0.3\\
$|f_{\rho\prime\prime}/f_\rho|$  &  0.006&0.002\\
$m_{\rho\prime\prime}$ [\mev]& 1790&20 \\
$\Gamma_{\rho\prime\prime}$ [\mev] & 290&40 \\     
$\phi_{\rho^{\prime\prime}}$ & 1.2&0.7\\

\end{tabular}

\label{tab:rho_r12}
\end{table}

%
%
\section{Cross-section}
\label{sec:cs}
The production cross-section, $\sigma_{V}$, for a vector meson, $V$, is determined in twelve bins of rapidity from 2.05 to 4.90
through
\begin{equation}
\frac{\deriv\sigma_V}{\deriv y}=
\frac{f_{\rm smog}N_{V}
}{
\Delta_y \, {\cal L}\, {\rm BF}_{V\rightarrow\pi\pi}
},
\end{equation}
where  $N_{V}$ is the number of vector mesons associated to the signal component $\cal{S}$, the integrated luminosity is represented by
${\cal L}$, the bin width by $\Delta_y$ and ${\rm BF}_{V\rightarrow\pi\pi}$ is the branching fraction of the decay of vector meson, $V$, to two charged pions.
For the $\rho$ meson, ${\rm BF}_{\rho\rightarrow\pi\pi}$ is taken to be $0.9901\pm0.0016$ by subtracting the branching fractions for the sub-dominant decay modes~\cite{PDG2024} from unity, while for the $\omega$ meson 
${\rm BF}_{\omega\rightarrow\pi\pi}=0.0153\pm0.0012$.
For the $\rho^\prime$ and $\rho^{\prime\prime}$ mesons, where the branching fractions are not measured, the
product ${\rm BF}_{V\rightarrow\pi\pi}(\deriv\sigma_V/{\deriv y})$ is reported.
A tiny correction, $f_{\rm smog} =0.9985\pm0.0001$ is 
 found through reconstruction of the longitudinal position of the dipion vertex, and accounts for vector meson photoproduction in fixed-target collisions that occur due to gas that was injected into the interaction region for a precision luminosity measurement~\cite{LHCb:2014vhh}.

The number of mesons in each rapidity bin is given by
\begin{equation}
N_V=\int_{2m_\pi}^{m_V+5\Gamma_V}{\cal S}(m) \deriv m \, ,
\end{equation}
with $\cal S$, depending on the model, defined in Eq.~\ref{eq:sod2},~\ref{eq:sodh1} or \ref{eq:best}, 
with the coefficients $f_\rho,f_\omega,f_{\rho\prime},f_{\rho\prime\prime},f_{\rm nr}$ set to zero except for the meson of interest.
In order to compare the cross-section results across experiments, the integration is performed, by convention, from threshold to $m_V+5\Gamma_{V}$, which are taken to be 1510, 826, 2950 and 3240\mev  for the $\rho,\omega,\rho^{\prime}$ and $\rho^{\prime\prime}$ meson, respectively.
The reference fits for the measurement of the $\rho$ and $\omega$ cross-section use the model of Eq.~\ref{eq:sodh1}.
Fitting with the model of Eq.~\ref{eq:sod2} increases the $\rho$ cross-section on average by 2\% but increases that of the $\omega$ meson by 34\%, showing the model-dependence of the results.
The fit with the model of Eq.~\ref{eq:best} increases the $\rho$ cross-section by 6\%, while that of the $\omega$ meson is unchanged. 
The total systematic uncertainty for the $\rho$ cross-section varies from 7 to 10\%, of which about 4\% is uncorrelated between bins.

To improve the stability of the $\omega$ cross-section as a function of rapidity, the fit range is constrained to be between 600 and 900\,\mev, and a correlated systematic uncertainty is assigned corresponding to the standard deviation of differences to the fit result performed between 400 and 1200\mev.  This is the dominant uncertainty.
The fits to extract the cross-section for the $\rho^\prime$ and $\rho^{\prime\prime}$ are performed in the mass range from 1.0 to 2.3\gev, with all parameters fixed except for $f_{\rho\prime}$ and $f_{\rho\prime\prime}$.  
The uncertainty includes a correlated
systematic component corresponding to the standard deviation of differences compared to the fit with all parameters free in the full mass range. Also in this case, it is the dominant uncertainty.

\begin{table}[t]
\begin{center}
\caption{Differential cross-section for the $\rho,\omega,\rho^{\prime}$ and $\rho^{\prime\prime}$ mesons in bins of rapidity, using the model of Eq.~\ref{eq:best}.  The first uncertainty includes statistical and uncorrelated systematics.  The second uncertainty represents the correlated systematics.}
\label{tab:cs}
\resizebox{1\columnwidth}{!}{
\begin{tabular}{c| r@{\:$\pm$\:}c@{\:$\pm$\:}l   | r@{\:$\pm$\:}c@{\:$\pm$\:}l   |  r@{\:$\pm$\:}c@{\:$\pm$\:}l   | r@{\:$\pm$\:}c@{\:$\pm$\:}l    }
Rapidity&
 \multicolumn{3}{c}{$[2.05,2.25]$}  &
  \multicolumn{3}{c}{$[2.25,2.50]$} & 
  \multicolumn{3}{c}{$[2.50,2.75]$}   & 
 \multicolumn{3}{c}{$[2.75,3.00]$}  
 \\
 \hline
$\frac{\deriv\sigma_\rho}{\deriv y}$[mb]&
473&23&40 &
412&16&33&
434&17&34&
453&18&36
\\
$\frac{\deriv\sigma_\omega}{\deriv y}$[mb]&
52&10&6 & 
24&4&3 & 
31&3&4 & 
32&4&4 
\\ 
${\rm BF}(\rho^\prime\rightarrow\pi\pi)\frac{\deriv\sigma_\rho\prime}{\deriv y}$[mb]&
1.55&0.12&0.33&
1.36&0.08&0.29&
1.46&0.07&0.31&
1.46&0.07&0.31
\\${\rm BF}(\rho^{\prime\prime}\rightarrow\pi\pi)\frac{\deriv \sigma_\rho\prime\prime}{\deriv y}$[mb]&
0.89&0.07&0.28&
1.04&0.05&0.32&
1.15&0.06&0.36&
1.26&0.06&0.39
\\
\multicolumn{5}{c}{ }  \\
   Rapidity&
\multicolumn{3}{c}{   $[3.00,3.25]$ } & 
\multicolumn{3}{c}{   $[3.25,3.50]$  }&
 \multicolumn{3}{c}{   $[3.50,3.75]$} & 
 \multicolumn{3}{c}{   $[3.75,4.00]$ } \\     
   \hline
$\frac{\deriv\sigma_\rho}{\deriv y}$[mb]&
464&18&37&
477&19&33& 
487&19&33& 493&19&33
\\ 
$\frac{\deriv\sigma_\omega}{\deriv y}$[mb]&
42&3&5 & 
45&3&5 &
45&3&5 &
36&3&4 
\\${\rm BF}(\rho^\prime\rightarrow\pi\pi)\frac{\deriv\sigma_\rho\prime}{\deriv y}$[mb]&
1.33&0.07&0.29&
1.18&0.59&0.25&
1.03&0.05&0.22&
0.93&0.05&0.20
\\
${\rm BF}(\rho^{\prime\prime}\rightarrow\pi\pi)\frac{\deriv\sigma_\rho\prime\prime}{\deriv y}$[mb]&
1.35&0.06&0.42&
1.22&0.05&0.38&
1.08&0.05&0.33&
0.92&0.05&0.28
\\
\multicolumn{5}{c}{ }  \\
   Rapidity&
  \multicolumn{3}{c}{  $[4.00,4.25]$ }& 
  \multicolumn{3}{c}{  $[4.25,4.50]$ }& 
  \multicolumn{3}{c}{  $[4.50,4.75]$ }& 
  \multicolumn{3}{c}{  $[4.75,4.90]$ }\\
\hline
$\frac{\deriv\sigma_\rho}{\deriv y}$[mb]&
498&20&28& 
520&21&30& 
556&22&38& 
625&43&44
\\ 
$\frac{\deriv\sigma_\omega}{\deriv y}$[mb]&
43&3&4 &
46&4&5 &
55&6&6 &
41&14&4 
\\
${\rm BF}(\rho^\prime\rightarrow\pi\pi)\frac{\deriv\sigma_\rho\prime}{\deriv y}$[mb]&
1.08&0.06&0.22&
1.29&0.06&0.27&
1.44&0.10&0.30&
1.56&0.29&0.33
\\
${\rm BF}(\rho^{\prime\prime}\rightarrow\pi\pi)\frac{\deriv\sigma_\rho\prime\prime}{\deriv y}$[mb]&
0.60&0.03&0.18&
0.56&0.03&0.17&
0.29&0.03&0.09&
0.02&0.04&0.01
\\
\end{tabular}
}
\end{center}

\end{table}

\begin{figure}[tb]
    \centering
        \includegraphics[scale=0.5]{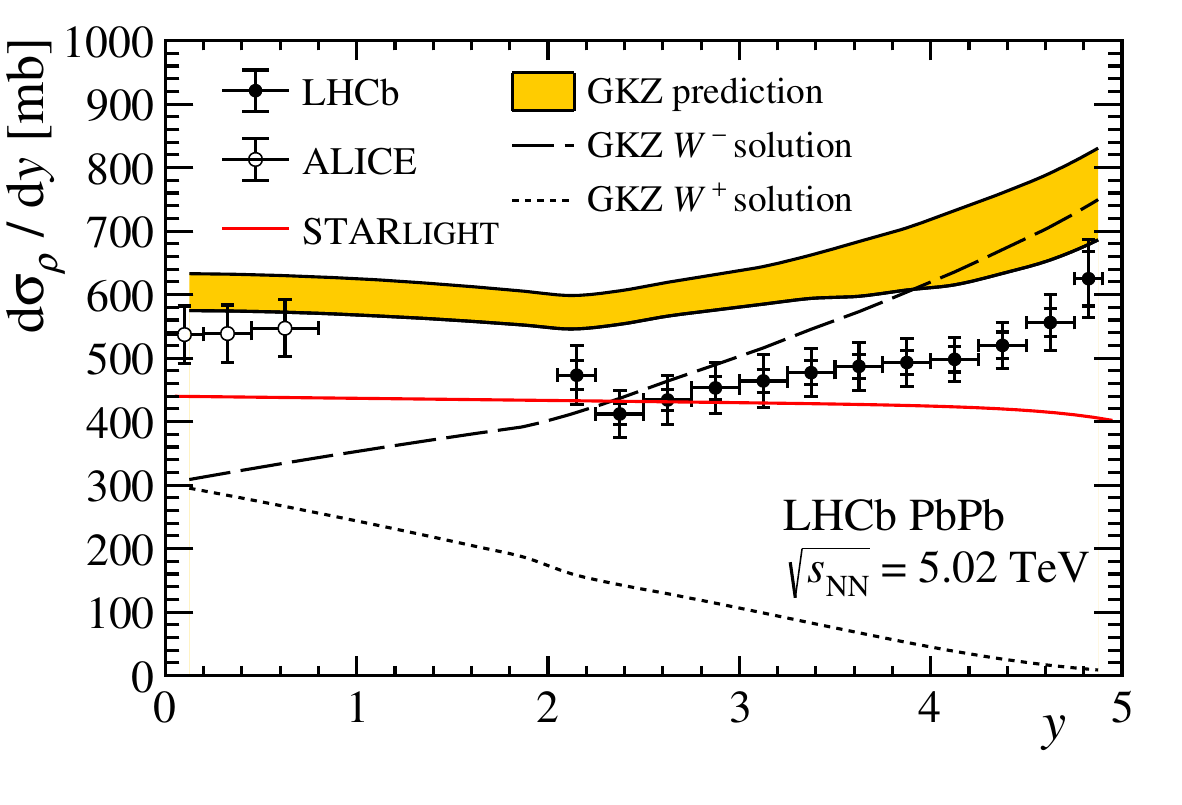}         
    \caption{Differential cross-section for coherent $\rho$ photoproduction as a function of rapidity. The LHCb points use the model of Eq.~\ref{eq:best} and the error bars represent the statistical and total uncertainties.  The shaded band corresponds to the estimated uncertainty on the GKZ prediction.
    The contributions due to the photon emitted from each ion are labelled as $W^+$ and $W^-$ and are explained further in Sec.~\ref{sec:supp}.
    Also shown are ALICE results at central rapidity and the output of the \textsc{STARlight} generator.
    }
    \label{fig:csrho}
\end{figure}

\begin{figure}[tb]
    \centering
        \includegraphics[scale=0.5]{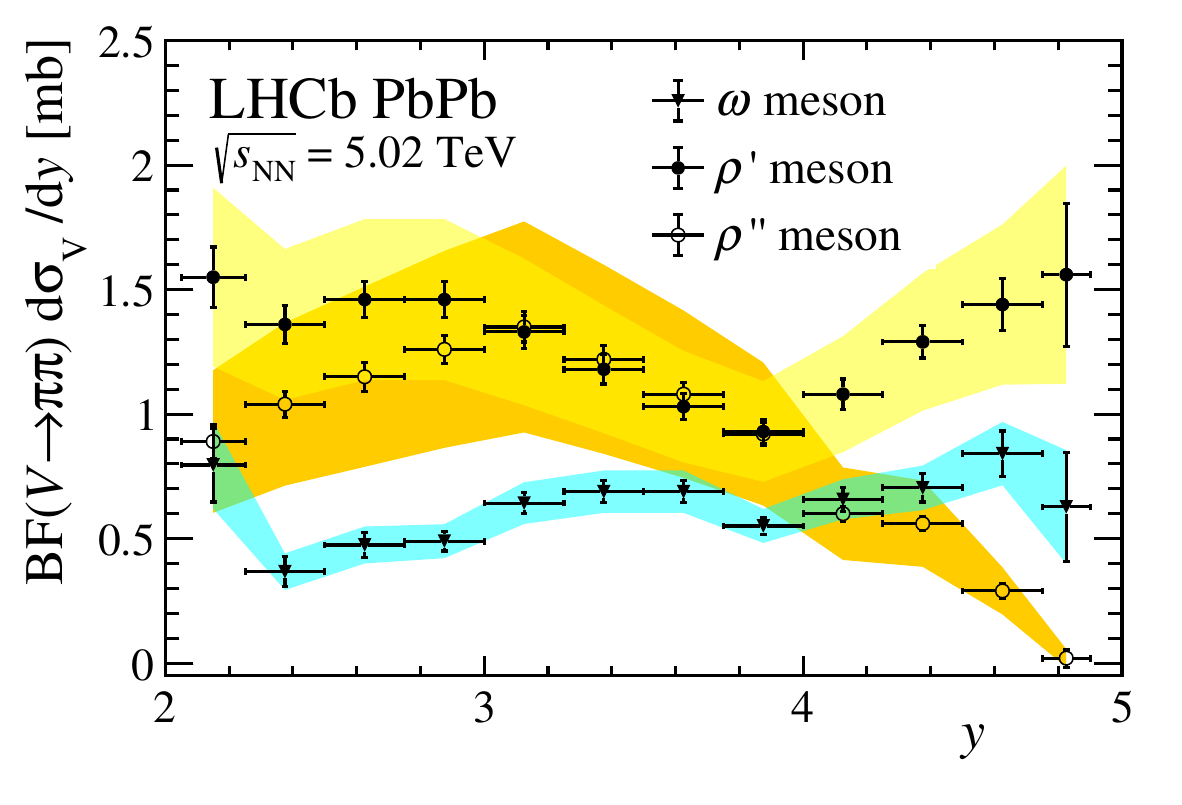}         
    \caption{Differential cross-section multiplied by the branching fraction to dipions, in bins of rapidity, for $\omega,\rho^\prime$ and $\rho^{\prime\prime}$ coherent photoproduction, calculated using the model of Eq.~\ref{eq:best}.
    The error bars represent the statistical uncertainty, while the bands correspond to the total uncertainty.
    }
    \label{fig:csomega}
\end{figure}

The results are given in Table~\ref{tab:cs} with the uncorrelated and correlated uncertainties quoted separately,
and are shown in Figs.~\ref{fig:csrho} and ~\ref{fig:csomega}.
For the $\rho$ meson, a comparison is made to ALICE data~\cite{ALICE:2020ugp}, to the predictions of Guzey, Kryshen and Zhalov (GKZ)~\cite{Guzey:2016piu}, and to the results of the \textsc{STARlight} generator~\cite{Klein:2016yzr}.
Further details of this calculation are provided in Sec.~\ref{sec:supp}.
The trend in the data for $\rho$ production is reproduced by the GKZ model, which takes account of elastic and inelastic nuclear shadowing.
However, the GKZ model predicts a slightly higher overall cross-section.
In other words, the data prefer more suppression than is given by the QCD effects considered in Ref.~\cite{Guzey:2016piu}.  
The measured cross-section roughly agrees with the \textsc{STARlight} generator although the $\rho$ measurement is systematically higher for rapidities above 3.5.
However, no strong conclusions can be drawn since \textsc{STARlight} simply uses a nuclear form factor to scale the HERA photoproduction results (see the comment in Sec.~4 of Ref.\cite{Frankfurt:2015cwa}).

The results for the $\omega$ photoproduction cross-section have a strong fit-model dependence with
the ratio $\sigma_\rho/\sigma_\omega$ varying between 
$9.8\pm1.0$ (Eq.~\ref{eq:sod2}),
$12.9\pm1.3$ (Eq.~\ref{eq:sodh1}) and 
\mbox{$13.5\pm1.4$} (Eq.~\ref{eq:best}).
These results are in agreement with the determinations of this ratio by the ZEUS~\cite{ZEUS:1996zse} and H1~\cite{H1:2020lzc} collaborations of $8.6\pm1.0$
and $10.3^{+3.3}_{-2.3}$, respectively, using
photoproduction on the proton at similar photon-proton centre-of-mass energies.
The ZEUS collaboration measured the $\omega$ cross-section directly in the channel $\omega\rightarrow\pip\pim\piz$, while H1 measured it through its interference, as performed here, in the channel $\omega\rightarrow\pip\pim$.

%
%
\section{Nuclear suppression factor}
\label{sec:supp}
Nuclear suppression effects are accessed by comparing the scaling between 
photoproduction on the proton, 
$\sigma_{\gamma p\rightarrow \rhoz p}$,
and photoproduction on the nucleus, 
$\sigma_{\gamma {\rm Pb}\rightarrow \rhoz {\rm Pb}}$.
First, the 
$\sigma_{\gamma p\rightarrow \rhoz p}$
results from the H1~\cite{H1:2020lzc} experiment are scaled to $\sigma_{\gamma {\rm Pb}\rightarrow \rhoz {\rm Pb}}$
using three models described below: the impulse approximation, a Glauber model, and a Gribov--Glauber model.  
Then, after accounting for the photon fluxes from  the lead nuclei, a prediction is obtained for the coherent production of $\rho$ mesons in PbPb collisions, which is compared to the experimental data.

The differential cross-section for ${\rm PbPb}\rightarrow {\rm Pb}\rho {\rm Pb}$ can be expressed~\cite{Guzey:2016piu} in terms of the photoproduction cross-section $\gamma {\rm Pb}\rightarrow \rho{\rm Pb}$ using the photon flux $\deriv N_\gamma/\deriv k$ for the number of photons $N_\gamma$ of energy $k$:
\begin{equation}
\frac{\deriv\sigma_{{\rm PbPb}\rightarrow {\rm Pb} \,\rho\, {\rm Pb}}}{\deriv y}
=
\biggl(k\frac{\deriv N_\gamma}{\deriv k}\biggr)^+\sigma_{\gamma {\rm Pb}\rightarrow \rho {\rm Pb}}(W^+)
+
\biggl(k\frac{\deriv N_\gamma}{\deriv k}\biggr)^-\sigma_{\gamma {\rm Pb}\rightarrow \rho {\rm Pb}}(W^-).
\label{eq:dsdy}
\end{equation}
The two terms on the right-hand side correspond to each of the ions being the photon emitter.
For LHCb, the `+' index corresponds to the higher-energy photon that is emitted by the ion moving with positive rapidity towards the detector.   
The photon energy, \mbox{$k^{\pm}=(m_\rho/2)\exp(\pm y)$}, and the centre-of-mass energy of the photon-ion system, \mbox{$W^\pm=(2k^\pm\sqrt{s})^{0.5}$}, are related to the fractional momentum of the partons in the nucleon by \mbox{$x=m_\rho^2/W^2$}.
Thus for PbPb UPC in LHCb, $W^+$ is in the range 170--710\gev corresponding to $x\sim [1,19]\times10^{-6}$ where Pomeron exchange dominates, while $W^-$ is in the range 5--22\gev, and $x\sim [0.001,0.020]$, where Reggeon exchange is also important.

\begin{figure}[tb]
    \centering
\advance\leftskip-2cm
\advance\rightskip-2cm
         \includegraphics[scale=0.4]{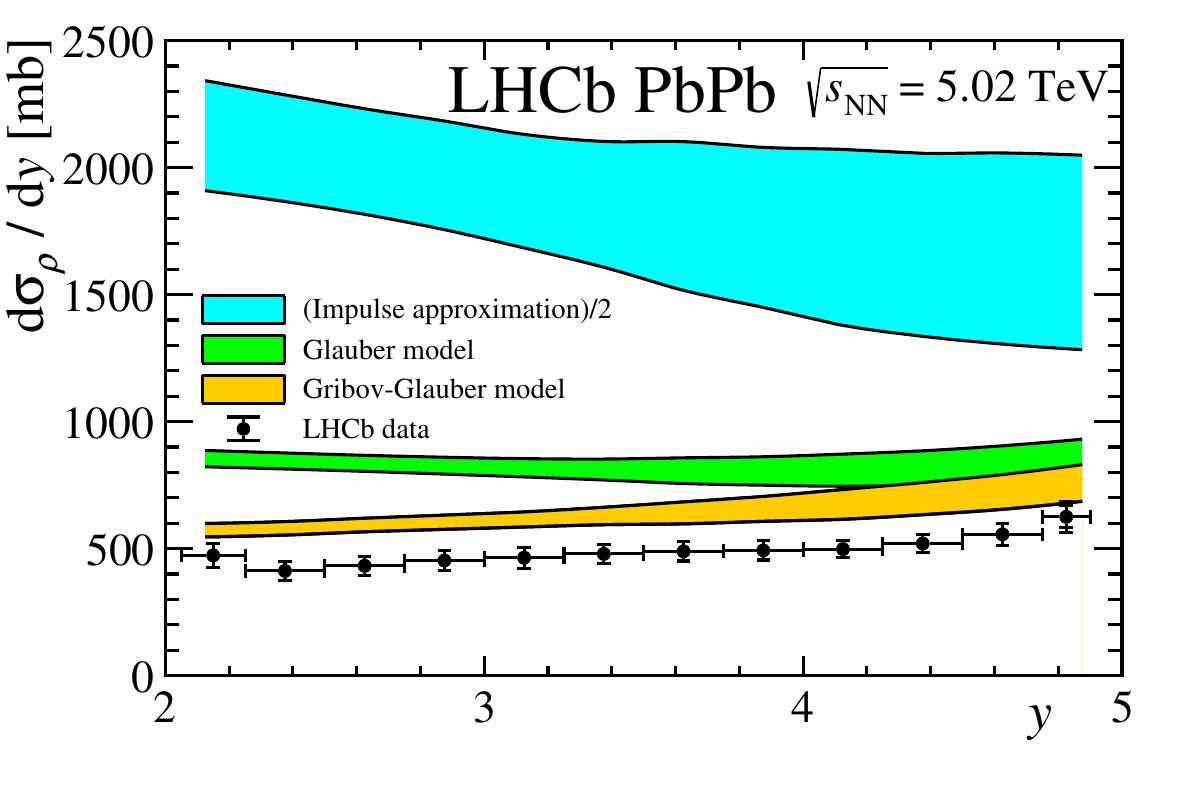} 
\includegraphics[scale=0.4]{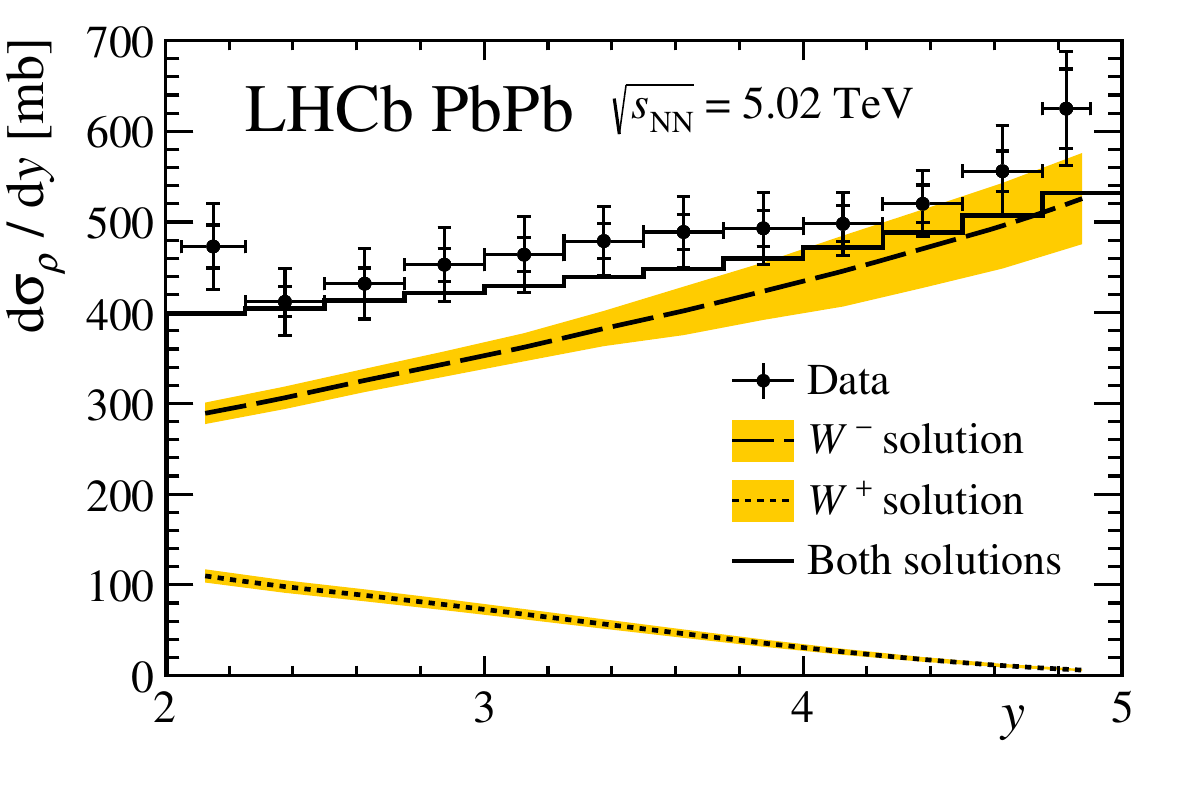}
    \caption{
    Differential cross-section for the $\rho$ meson as a function of rapidity for data (left) compared to the impulse approximation, Glauber model, and Gribov--Glauber model and (right) with
    the estimated contributions from photoproduction on each ion in the 
    Gribov--Glauber model after including nuclear suppression factors.
    The impulse approximation has been scaled down by a factor of two so it can be viewed with this axis-range.
    \label{fig:3sol}}
\end{figure}

In the absence of nuclear effects, the $\rho$ photoproduction cross-section on the ion scales
from the cross-section on the proton according to the number of nucleons and is given by the impulse approximation~\cite{Chew:1952fca}.
However, due to the large cross-section for $\rho$-nucleon scattering, as the meson traverses the nucleus, multiple interactions are possible.
This can be taken into account in the Glauber model that considers elastic scatters, or
extended to include inelastic scatters with a Gribov--Glauber model~\cite{Frankfurt:2015cwa}.
Details of these calculations are given in the Appendix. 
The predictions for the impulse approximation, Glauber and Gribov--Glauber models are shown in the left panel of Fig.~\ref{fig:3sol}, where it can be seen that the elastic and inelastic rescattering effects have a major impact on the result.
The data are clearly inconsistent with the impulse approximation, while nuclear shadowing alone is also insufficient to provide a complete description.  
Including inelastic scattering improves the overall picture but some additional suppression is needed. 

Equation~\ref{eq:dsdy} shows there is 
a two-fold ambiguity in determining the photon emitter.
However, at the highest rapidities, irrespective of nuclear suppression, the backward-going lead nucleus is almost always the photon emitter.
As a result, the first term on the right-hand side only contributes 3\% in the rapidity bin from 4.75 to 4.90, well within the systematic uncertainties.
Ignoring this `+' solution, Eq.~\ref{eq:dsdy} predicts a $\rho$ cross-section for PbPb collisions without nuclear suppression of $3330\pm770$~mb,
to be compared with the measured value of $625\pm43\pm44$~mb.
In the perturbative regime where the Pomeron is equivalent to two gluons, the nuclear suppression factor has been defined~\cite{Guzey:2020ntc} with respect to the impulse approximation using
\begin{equation}    S_{\rm IA}=\sqrt{
\frac{\sigma_{\gamma \rm Pb\rightarrow\rho Pb}}
{\sigma^{\rm IA}_{\gamma \rm Pb\rightarrow\rho Pb}}},
\label{eq:iasup}
\end{equation}
since for perturbatively produced objects like the $\jpsi$ meson, this quantity is related to the ratio of the gluon parton distribution functions in the nucleus and the proton.
The same definition of $S$ is used here, although the relationship with the parton distribution functions is not possible using the nonperturbatively produced $\rho$ meson. 
A suppression factor of $S_{\rm IA}=0.43\pm0.05$ is measured at a centre-of-mass energy $W=5.4\gev$ or equivalently, a fractional parton momentum of $x=0.018$.

The LHCb data also have sensitivity to the nuclear suppression factor at high $W$ (low $x$), since at low rapidities the `+' solution contributes about one third.
A fit to all the data is performed as follows.
Equation~\ref{eq:dsdy} is modified to include a nuclear suppression factor,
\begin{equation}
\label{eq:pbpbwiths}
\frac{\deriv\sigma_{{\rm PbPb}\rightarrow {\rm Pb} \,\rho\, {\rm Pb}}}{\deriv y}
=
(S^+_{\rm model})^2
\biggl(k\frac{\deriv N_\gamma}{\deriv k}\biggr)^+\sigma^{\rm model}_{\gamma {\rm Pb}\rightarrow \rho {\rm Pb}}(W^+)
+
(S^-_{\rm model})^2
\biggl(k\frac{\deriv N_\gamma}{\deriv k}\biggr)^-\sigma^{\rm model}_{\gamma {\rm Pb}\rightarrow \rho {\rm Pb}}(W^-),
\end{equation}
where one suppression factor is used for high and one for low $W$, 
and
\begin{equation}    S_{\rm model}=\sqrt{
\frac{\sigma_{\gamma \rm Pb\rightarrow\rho Pb}}
{\sigma^{\rm model}_{\gamma \rm Pb\rightarrow\rho Pb}}},
\label{eq:s}
\end{equation} is more-generally
defined with respect to the impulse approximation, Glauber, or Gribov-Glauber model.

A chisquare fit is performed with a Lagrange multiplier to take account of the correlated uncertainties, assuming that the uncertainties  
on the theory predictions are all correlated.
The quantity that is minimised is
\begin{equation}
\chi^2(S^+,S^-,\lambda)=
\sum_i \biggl(\frac{x^{\rm exp}_i-x^{\rm th}_i}{\sigma^{\rm uncor}_i}\biggr)^2+\lambda^2\,,
\end{equation}
where 
$x^{\rm exp}$ are the measured values in Table~\ref{tab:cs},
$\sigma_i^{\rm uncor}$ are the uncorrelated experimental uncertainties,
$x^{\rm th}_i=
{\deriv\sigma_{\rm PbPb\rightarrow Pb \rho Pb}/ \deriv y}+\lambda\sigma_i^{\rm cor}$
with 
${\deriv\sigma_{\rm PbPb\rightarrow Pb \rho Pb}/ \deriv y}$ given by Eq.~\ref{eq:pbpbwiths},
and $\sigma_i^{\rm cor}$ the correlated uncertainties associated with the experimental uncertainties, the photon flux, the parametrisation of $\sigma_{\rho N}$, and 
the dispersion of $P(\sigma)$
(as described in the Appendix).

With the suppression factor measured relative to the impulse approximation, the fit
returns values of $S_{\rm IA}^-=0.39\pm0.03$ and $S_{\rm IA}^+=0.05\pm0.11$ while $\lambda=0.6\pm0.6$.
This result shows that the events produced by higher-energy photons (lower-energy gluons) experience more nuclear suppression than those produced by lower-energy photons (higher energy gluons), as predicted by the Glauber and Gribov--Glauber calculations.
A similar effect, albeit at a much reduced level in the pertubative \jpsi process, has been reported in Refs.~\cite{Guzey:2020ntc, Mantysaari:2023xcu} using ALICE and CMS UPC data~\cite{ALICE:2023jgu,CMS:2023snh}.
To take account of rescattering contributions, the nuclear suppression is calculated with respect to the Glauber approximation that includes elastic  contributions predicted by QCD.
This returns
\mbox{$S_{\rm Glauber}^-=0.78\pm0.04$} and $S_{\rm Glauber}^+=0.30\pm0.11$ while $\lambda=0.9\pm0.9$.
This shows that nonlinear effects are much more significant for the low-$x$ gluons.
Finally, after including inelastic contributions using the Gribov--Glauber model, the relative suppression factors are
\mbox{$S_{\rm Gribov-Glauber}^-=0.84\pm0.04$} and $S^+_{\rm Gribov-Glauber}=0.83\pm0.06$ while $\lambda=0.7\pm0.9$.
The result is shown in the right panel of Fig.~\ref{fig:3sol}.
The data is in closer agreement to the model although some additional suppression is still needed, which could be due to saturation effects or an  increased dispersion of $P(\sigma)$.

\section{Conclusions}
\label{sec:conclude}

Using a clean sample of dipions produced in ultraperipheral PbPb collisions, the
invariant-mass distribution shows a prominent $\rhoz$ peak, but also contributions from nonresonant production, the $\omega$ resonance and higher-mass resonances.  
Hints of a resonance at 1.7\gev reported by the ALICE and STAR collaborations are unambiguously confirmed with the larger LHCb data sample that, in the model chosen, also requires a resonance at 1.35\gev.
The cross-sections for coherent photoproduction of $\rhoz,\omega,\rho^{\prime}$ and $\rho^{\prime\prime}$ mesons are measured as functions of rapidity.
The coherent $\omega$ cross-section is found to be 
about a factor ten lower than that of the $\rhoz$ meson in agreement with photoproduction measurements at HERA.

The differential cross-section for the $\rhoz$ meson shows a slow rise with rapidity, in agreement with the  GKZ model that includes linear and nonlinear QCD effects using a Gribov--Glauber calculation to describe multiple nucleon interactions.
The observed cross-section is lower by about 30\%, indicating 
that additional suppression is required.  
In PbPb interactions, there is a two-fold ambiguity in determining the photon emitter. 
However, by utilising the different cross-section dependencies with rapidity, there is sensitivity to both high and low $x$-values.
Compared to the Gribov--Glauber model, suppression factors of $0.84\pm0.04$ for values of $x\sim 5\times 10^{-3}$ and of $0.83\pm0.06$ for $x\sim 5\times 10^{-6}$ are obtained.  
Additional suppression is observed and this could be caused by saturation effects.
However, 
models that explicitly include saturation cannot be used to predict the cross-section since the $\rhoz$ meson is too light to apply perturbation theory.
We thus rely on a phenomenological model that scales and extrapolates fits to fixed-target and H1 data for photoproduction on the proton. 
The LHCb photon-nucleon centre-of-mass energies in the range 170--760\gev require an extrapolation of the H1 data, while energies in the range 5--23\gev are
only measured in fixed-target experiments, are less precise, and are also in a region where Reggeon exchange is important.
Thus, new measurements of photoproduction on the proton are highly desirable at these energies and can be obtained at forward rapidities in proton-lead collisions at the LHC, or using fixed-target gaseous collisions at LHCb.

\section*{Appendix}\label{appendix}
Using the photon flux, Eq.~\ref{eq:dsdy} relates the 
differential cross-section for ${\rm PbPb}\rightarrow {\rm Pb}\rho {\rm Pb}$ to the photoproduction cross-section $\gamma {\rm Pb}\rightarrow \rho{\rm Pb}$, expressed using the impulse approximation, the Glauber model, or the Gribov--Glauber model.

The photon flux for quasi-real photons can be calculated in the equivalent photon approximation~\cite{Baur:2001jj} as a function of $k$ and impact parameter $b$.  Restricting $b$ to be greater than twice the radius of the ions, as is required in UPC, gives
\begin{equation}
\label{eq:flux}
    k\frac{\deriv N_\gamma}{\deriv k}
    =
    \frac{2Z^2\alpha}{\pi}
    \biggl(\zeta K_0(\zeta)K_1(\zeta)-\frac{\zeta^2}{2}[K_1^2(\zeta)-K^2_0(\zeta)]\biggr),
\end{equation}
with $\zeta=2R_{\rm Pb}k/\upgamma$, where $R_{\rm Pb}=6.68\fm$~\cite{Jones:2014aoa} is the radius of the ion, $\upgamma$ is the nucleus Lorentz factor, $Z$ is the charge of the ion, $\alpha$ is the fine-structure constant and $K_0,K_1$ are modified Bessel functions.

The impulse approximation is given 
by
\begin{equation}
\sigma^{\rm IA}_{\gamma {\rm Pb}\rightarrow\rho {\rm Pb}}(W)=
\frac{\deriv \sigma_{\gamma p\rightarrow \rho p}}{\deriv t}
(W,t=0)
\int_{-\infty}^{t_{\rm min}} |F_{\rm Pb}(t)|^2\deriv t,
\label{eq:ia}
\end{equation}
which only takes into account the nuclear form factor, $F_{\rm Pb}(t)$, and neglects all other potential nuclear effects.
The four-momentum-transfer squared is represented by $t\approx-\pt^2$ with $t_{\rm min}=-m_\rho^4m_p^2/{W^4}$.
The form factor is a Fourier transform of the nuclear density,
\begin{equation}
F_{\rm Pb}(q)=\frac{4\pi}{q}\int_0^\infty r\sin(qr)\uprho(r)\deriv r,
\end{equation}
where the momentum transfer $q\approx\pt$ and
$\uprho(r)=\uprho_p(r)+\uprho_n(r)$ with $\uprho_p$ and $\uprho_n$ being the density of protons and neutrons in the lead ion that are usually described through a Woods--Saxon distribution
\begin{equation}
\uprho(r)=\frac{\uprho_0}{1+\exp((r-R)/d)}\, ,
\end{equation}
with $R=6.68 \, (6.67)\fm$ for the radius and
$d=0.45 \, (0.55) \fm$ for the skin depth of protons (neutrons)~\cite{Tarbert:2013jze,Jones:2014aoa}.
Requiring that $\int\uprho_p\deriv V=82$
and $\int\uprho_n\deriv V=208-82$ gives $\uprho_0=0.063 \, (0.095)\fm^{-3}$ for protons (neutrons).

The H1 collaboration measured $\rho$ production in the energy range $20<W<80\gev$, while the
range considered here is larger.
For $W< 20\gev$, Reggeon as well as Pomeron contributions are important, while for $W>80\gev$ an extrapolation of the HERA data is required.
By fitting their data combined with fixed target data, the H1 collaboration found~\cite{H1:2020lzc}
\begin{equation}
\label{eq:h1par}
    \sigma_{\gamma p\rightarrow\rho p} (W)=
    \sigma(W_0)
    \biggl\{
    \biggl(
    \frac{W}{ W_0}\biggr)^{\delta_{\cal P}}
    + f_{\cal R} 
    \biggl(\frac{W}{W_0}\biggr)^{\delta_{\cal R}}
    \biggr\}\, ,
\end{equation}
with a cross-section of $\sigma(W_0)=10.66\pm0.11^{+0.75}_{-1.00}\mub$ at $W_0=40\gev$,
while the Pomeron and Reggeon powers are $\delta_{\cal P}=0.207\pm0.015^{+0.053}_{-0.033}$ and 
$\delta_{\cal R}=-1.45\pm0.12^{+0.35}_{-0.21}$,
and the fraction of Reggeon contribution, $f_{\cal R}=0.020\pm0.007^{+0.029}_{-0.013}$.
The $t$-dependence for their data is approximately exponential: $\frac{\deriv\sigma}{\deriv t}=\frac{\deriv\sigma}{\deriv t}|_{t=0}\exp(bt)$,
which by integration yields 
$\frac{\deriv\sigma}{\deriv t}|_{t=0}=b
\sigma_{\gamma p\rightarrow\rho p}$.
The $b$-dependence was also determined with H1 and fixed-target data using just a single equation 
of the form
\begin{equation}
    b(W)=b(W_0)+4\alpha_1\log(W/W_0)\, ,
    \label{eq:beq}
\end{equation}
that describes the data well in both the Reggeon- and Pomeron-dominated regions.
The shrinkage of the diffraction peak was observed with fitted
values of \mbox{$b(W_0)=9.59\pm0.14^{+0.14}_{-0.12}\gev^{-2}$}
and $\alpha_1=0.233\pm0.064^{+0.020}_{-0.038}\gev^{-2}$.
Combining these results gives
\begin{equation}
    \frac{\deriv\sigma_{\gamma p\rightarrow\rho p}}{\deriv t}(W,t=0)
    =\biggl(b(W_0)+4\alpha_1\log(W/W_0)\biggr)
    \sigma(W_0)
    \biggl\{
    \biggl(
    \frac{W}{ W_0}\biggr)^{\delta_{\cal P}}
    + f_{\cal R} 
    \biggl(\frac{W}{ W_0}\biggr)^{\delta_{\cal R}}
    \biggr\}.
    \label{eq:dsdt0}
\end{equation}
Using this expression in 
Eq.~\ref{eq:ia} and Eq.~\ref{eq:dsdy} allows a calculation of the differential cross-section for UPC $\rho$ production  in the impulse approximation.

Multiple interactions are taken into account using a Glauber calculation, assuming vector-meson dominance and using the optical theorem to relate the total cross-section to the imaginary part of the forward scattering amplitude.
Following Ref.~\cite{Frankfurt:2015cwa},
the photoproduction cross-section on the nucleus can be written,
\begin{equation}
\sigma^{\rm Glauber}_{\gamma{\rm Pb}\rightarrow \rho{\rm Pb}}=
\biggl(\frac{e}{f_\rho}\biggr)^2 
\int \deriv^2\vec b \,
\biggl|
1-e^{-\sigma_{\rho N}T_A(b)/2}
\biggr|^2 \,,
\label{eq:glauber}
\end{equation}
where $T_A$ is the thickness of the nucleus at impact parameter $\vec{b}$ and $\sigma_{\rho N}$ is the total $\rho$--nucleon cross-section.
The experimentally derived quantity $f_\rho^2/(4\pi)=2.01\pm0.01$ is used for the $\rho$--$\gamma$ coupling constant.
The thickness is found from the 
nuclear density,
\begin{equation}
T_A(b)=\int_{-\infty}^{+\infty} \deriv z
\biggl(
\uprho_p(\sqrt{b^2+z^2})+
\uprho_n(\sqrt{b^2+z^2})\biggr).
\end{equation}
The total cross-section for $\rho$--nucleon scattering can be found
once again from the optical theorem,
\begin{equation}
   \frac{\deriv\sigma_{\gamma p\rightarrow\rho p}}{\deriv t}(W,t=0)=
   \frac{1}{16\pi} \biggl(\frac{e}{f_\rho}\biggr)^2\sigma^2_{\rho N}(W),
\label{eq:totfromphoto}
\end{equation}
with $\deriv\sigma_{\gamma p\rightarrow\rho p}/{\deriv t}$
given by Eq.~\ref{eq:dsdt0}.

The Glauber approximation only considers elastic scatters and needs to be extended to include inelastic scatters. 
This can be done with a Gribov--Glauber model as described in Ref.~\cite{Frankfurt:2015cwa}, and 
Eq.~\ref{eq:glauber} is modified to 
\begin{equation}
\sigma^{\rm Gribov-Glauber}_{\gamma{\rm Pb}\rightarrow \rho{\rm Pb}}=
\biggl(\frac{e}{f_\rho}\biggr)^2
\int \deriv^2\vec b\,
\biggl|
\int
\deriv\sigma P(\sigma)
\biggl(1-e^{-\sigma T_A(b)/2}\bigg)
\biggr|^2,
\label{eq:gribov}
\end{equation}
where now instead of a single value for $\sigma_{\rho N}$, a distribution, $P(\sigma)$, of $\rho$-nucleon cross-sections is used with
\begin{equation}
P(\sigma)=C\frac{1}{1+(\sigma/\sigma_0)^2}e^{-(\sigma/\sigma_0-1)^2/\Omega^2}.
\end{equation}
The parameters, $C,\sigma_0,\Omega$ are
detailed in Ref.~\cite{Frankfurt:2015cwa}, where $C$ is a normalisation constant, the average total cross-section $\int\sigma P(\sigma)\deriv\sigma=\sigma_{\rho N}$,
and the width of the distribution, $\Omega$ is constrained by data.

The uncertainties on the predictions are calculated as the quadratic sum of the uncertainty on the photon flux, obtained by changing the ion radius by 0.5\fm, and the uncertainty on 
$\deriv\sigma_{\gamma p\rightarrow p}/\deriv t(W,t=0)$, calculated from the uncertainties and correlations on the parameters of Eq.~\ref{eq:dsdt0} given in Ref.~\cite{H1:2020lzc}.
For the Gribov--Glauber model, an additional uncertainty is added in quadrature, obtained by changing the dispersion of $P(\sigma)$ as described in Ref.~\cite{Frankfurt:2015cwa}.

%% file: acknowledgements.tex
\section*{Acknowledgements}
%
%
\noindent We express our gratitude to our colleagues in the CERN
accelerator departments for the excellent performance of the LHC. We
thank the technical and administrative staff at the LHCb
institutes.
We acknowledge support from CERN and from the national agencies:
ARC (Australia);
CAPES, CNPq, FAPERJ and FINEP (Brazil); 
MOST and NSFC (China); 
CNRS/IN2P3 (France); 
BMBF, DFG and MPG (Germany); 
INFN (Italy); 
NWO (Netherlands); 
MNiSW and NCN (Poland); 
MCID/IFA (Romania); 
MICIU and AEI (Spain);
SNSF and SER (Switzerland); 
NASU (Ukraine); 
STFC (United Kingdom); 
DOE NP and NSF (USA).
We acknowledge the computing resources that are provided by ARDC (Australia), 
CBPF (Brazil),
CERN, 
IHEP and LZU (China),
IN2P3 (France), 
KIT and DESY (Germany), 
INFN (Italy), 
SURF (Netherlands),
Polish WLCG (Poland),
IFIN-HH (Romania), 
PIC (Spain), CSCS (Switzerland), 
and GridPP (United Kingdom).
We are indebted to the communities behind the multiple open-source
software packages on which we depend.
Individual groups or members have received support from
Key Research Program of Frontier Sciences of CAS, CAS PIFI, CAS CCEPP, 
Fundamental Research Funds for the Central Universities,  and Sci.\ \& Tech.\ Program of Guangzhou (China);
Minciencias (Colombia);
EPLANET, Marie Sk\l{}odowska-Curie Actions, ERC and NextGenerationEU (European Union);
A*MIDEX, ANR, IPhU and Labex P2IO, and R\'{e}gion Auvergne-Rh\^{o}ne-Alpes (France);
Alexander-von-Humboldt Foundation (Germany);
ICSC (Italy); 
Severo Ochoa and Mar\'ia de Maeztu Units of Excellence, GVA, XuntaGal, GENCAT, InTalent-Inditex and Prog.~Atracci\'on Talento CM (Spain);
SRC (Sweden);
the Leverhulme Trust, the Royal Society and UKRI (United Kingdom).

We thank Misha Ryskin for many helpful discussions and  Vadim Guzey for his assistance with the calculation of the nuclear suppression
factors and for making his results available to us.

%% file: Authorship_LHCb-PAPER-2024-042.tex
\centerline
{\large\bf LHCb collaboration}
\begin
{flushleft}
\small
R.~Aaij$^{38}$\lhcborcid{0000-0003-0533-1952},
A.S.W.~Abdelmotteleb$^{57}$\lhcborcid{0000-0001-7905-0542},
C.~Abellan~Beteta$^{51}$\lhcborcid{0009-0009-0869-6798},
F.~Abudin{\'e}n$^{57}$\lhcborcid{0000-0002-6737-3528},
T.~Ackernley$^{61}$\lhcborcid{0000-0002-5951-3498},
A. A. ~Adefisoye$^{69}$\lhcborcid{0000-0003-2448-1550},
B.~Adeva$^{47}$\lhcborcid{0000-0001-9756-3712},
M.~Adinolfi$^{55}$\lhcborcid{0000-0002-1326-1264},
P.~Adlarson$^{82}$\lhcborcid{0000-0001-6280-3851},
C.~Agapopoulou$^{14}$\lhcborcid{0000-0002-2368-0147}
% ...

{\footnotesize \it

% ...
$^{14}$Universit{\'e} Paris-Saclay, CNRS/IN2P3, IJCLab, Orsay, France\\
% ...
$^{38}$Nikhef National Institute for Subatomic Physics, Amsterdam, Netherlands\\
% ...
$^{47}$Instituto Galego de F{\'\i}sica de Altas Enerx{\'\i}as (IGFAE), Universidade de Santiago de Compostela, Santiago de Compostela, Spain\\
% ...
$^{51}$Physik-Institut, Universit{\"a}t Z{\"u}rich, Z{\"u}rich, Switzerland\\
% ...
$^{55}$H.H. Wills Physics Laboratory, University of Bristol, Bristol, United Kingdom\\
% ...
$^{57}$Department of Physics, University of Warwick, Coventry, United Kingdom\\
% ...
$^{60}$School of Physics and Astronomy, University of Glasgow, Glasgow, United Kingdom\\
$^{61}$Oliver Lodge Laboratory, University of Liverpool, Liverpool, United Kingdom\\
% ...
$^{69}$Syracuse University, Syracuse, NY, United States\\
% ...
$^{82}$Department of Physics and Astronomy, Uppsala University, Uppsala, Sweden, associated to $^{60}$\\
% ...
}
\end{flushleft}

%% file: LHCb-PAPER.bib
@article{LHCb-PAPER-2022-012,
      author         = "Aaij, R. and others",
      title          = "{Study of exclusive photoproduction of charmonium in ultra-peripheral lead-lead collisions}",
      collaboration  = "LHCb collaboration",
      report         = "{LHCb-PAPER-2022-012, CERN-EP-2022-108}",
      eprint         = "2206.08221",
      archivePrefix  = "arXiv",
      primaryClass   = "hep-ex",
      year           = "2023",
      journal        = "JHEP",
      volume         = "06",
      pages          = "146",
      doi            = "10.1007/JHEP06(2023)146",
}


%% file: main.bib
@article{Guzey:2016piu,
    author = "Guzey, V. and Kryshen, E. and Zhalov, M.",
    title = "Coherent photoproduction of vector mesons in ultraperipheral heavy ion collisions: Update for \uppercase{r}un 2 at the \uppercase{CERN} \uppercase{L}arge \uppercase{H}adron \uppercase{C}ollider",
    eprint = "1602.01456",
    archivePrefix = "arXiv",
    primaryClass = "nucl-th",
    doi = "10.1103/PhysRevC.93.055206",
    journal = "Phys. Rev.",
    volume = "C93",
    number = "5",
    pages = "055206",
    year = "2016"
}

@article{H1:2020lzc,
    author = "Andreev, V. and others",
    collaboration ="H1 collaboration",
    title = "Measurement of exclusive $\pi^{+}\pi^{-}$ and $\rho^0$ meson photoproduction at \uppercase{HERA}",
    eprint = "2005.14471",
    archivePrefix = "arXiv",
    primaryClass = "hep-ex",
    reportNumber = "DESY-20-080",
    doi = "10.1140/epjc/s10052-020-08587-3",
    journal = "Eur. Phys. J.",
    volume = "C80",
    number = "12",
    pages = "1189",
    year = "2020"
}

@article{Glauber:1970jm,
    author = "Glauber, R. J. and Matthiae, G.",
    title = "High-energy scattering of protons by nuclei",
    doi = "10.1016/0550-3213(70)90511-0",
    journal = "Nucl. Phys.",
    volume = "B21",
    pages = "135--157",
    year = "1970"
}

@article{Accardi:2012qut,
    author = "Accardi, A. and others",
    editor = "Deshpande, A. and Meziani, Z. E. and Qiu, J. W.",
    title = "Electron \uppercase{I}on \uppercase{C}ollider: The next \uppercase{QCD} frontier: Understanding the glue that binds us all",
    eprint = "1212.1701",
    archivePrefix = "arXiv",
    primaryClass = "nucl-ex",
    reportNumber = "BNL-98815-2012-JA, JLAB-PHY-12-1652",
    doi = "10.1140/epja/i2016-16268-9",
    journal = "Eur. Phys. J.",
    volume = "A52",
    number = "9",
    pages = "268",
    year = "2016"
}

@article{STAR:2007elq,
    author = "Abelev, B. I. and others",
    collaboration ="STAR collaboration",
    title = "$\rho^0$ photoproduction in ultraperipheral relativistic heavy ion collisions at $\sqrt{s_{NN}}$ = 200 \uppercase{G}e\uppercase{V}",
    eprint = "0712.3320",
    archivePrefix = "arXiv",
    primaryClass = "nucl-ex",
    doi = "10.1103/PhysRevC.77.034910",
    journal = "Phys. Rev.",
    volume = "C77",
    pages = "034910",
    year = "2008"
}

@article{STAR:2017enh,
    author = "Adamczyk, L. and others",
    collaboration ="STAR collaboration",
    title = "Coherent diffractive photoproduction of $\rho^{0}$mesons on gold nuclei at 200 \uppercase{G}e\uppercase{V}/nucleon-pair at the Relativistic Heavy Ion Collider",
    eprint = "1702.07705",
    archivePrefix = "arXiv",
    primaryClass = "nucl-ex",
    doi = "10.1103/PhysRevC.96.054904",
    journal = "Phys. Rev.",
    volume = "C96",
    number = "5",
    pages = "054904",
    year = "2017"
}

@article{STAR:2011wtm,
    author = "Agakishiev, G. and others",
    collaboration ="STAR collaboration",
    title = "$\rho^{0}$ photoproduction in \uppercase{A}u\uppercase{A}u collisions at $\sqrt{s_{NN}}$=62.4 \uppercase{G}e\uppercase{V} with \uppercase{STAR}",
    eprint = "1107.4630",
    archivePrefix = "arXiv",
    primaryClass = "nucl-ex",
    doi = "10.1103/PhysRevC.85.014910",
    journal = "Phys. Rev.",
    volume = "C85",
    pages = "014910",
    year = "2012"
}

@article{ALICE:2021jnv,
    author = "Acharya, Shreyasi and others",
    collaboration = "ALICE colaboration",
    title = "First measurement of coherent $\rho^0$ photoproduction in ultra-peripheral \uppercase{X}e\textendash{}\uppercase{X}e collisions at $\sqsnn=5.44$ \uppercase{T}e\uppercase{V}",
    eprint = "2101.02581",
    archivePrefix = "arXiv",
    primaryClass = "nucl-ex",
    reportNumber = "CERN-EP-2020-252",
    doi = "10.1016/j.physletb.2021.136481",
    journal = "Phys. Lett.",
    volume = "B820",
    pages = "136481",
    year = "2021"
}

@article{ALICE:2020ugp,
    author = "Acharya, Shreyasi and others",
    collaboration = "ALICE colaboration",
    title = "Coherent photoproduction of $\rho^{0}$ vector mesons in ultra-peripheral \uppercase{P}b--\uppercase{P}b collisions at $ \sqsnn = 5.02$ \uppercase{T}e\uppercase{V}",
    eprint = "2002.10897",
    archivePrefix = "arXiv",
    primaryClass = "nucl-ex",
    reportNumber = "CERN-EP-2020-021",
    doi = "10.1007/JHEP06(2020)035",
    journal = "JHEP",
    volume = "06",
    pages = "035",
    year = "2020"
}

@article{Klein:2016yzr,
    author = "Klein, Spencer R. and Nystrand, Joakim and Seger, Janet and Gorbunov, Yuri and Butterworth, Joey",
    title = "STARlight: A \uppercase{M}onte \uppercase{C}arlo simulation program for ultra-peripheral collisions of relativistic ions",
    eprint = "1607.03838",
    archivePrefix = "arXiv",
    primaryClass = "hep-ph",
    doi = "10.1016/j.cpc.2016.10.016",
    journal = "Comput. Phys. Commun.",
    volume = "212",
    pages = "258--268",
    year = "2017"
}

@article{ZEUS:1997rof,
    author = "Breitweg, J. and others",
    collaboration ="ZEUS collaboration",
    title = "Elastic and proton dissociative $\rho^0$ photoproduction at \uppercase{HERA}",
    eprint = "hep-ex/9712020",
    archivePrefix = "arXiv",
    reportNumber = "DESY-97-237",
    doi = "10.1007/s100520050136",
    journal = "Eur. Phys. J.",
    volume = "C2",
    pages = "247--267",
    year = "1998"
}

@article{H1:2009cml,
    author = "Aaron, F. D. and others",
    collaboration ="H1 collaboration",
    title = "Diffractive electroproduction of $\rho$ and $\phi$ mesons at \uppercase{HERA}",
    eprint = "0910.5831",
    archivePrefix = "arXiv",
    primaryClass = "hep-ex",
    reportNumber = "DESY-09-093",
    doi = "10.1007/JHEP05(2010)032",
    journal = "JHEP",
    volume = "05",
    pages = "032",
    year = "2010"
}

@article{H1:1996prv,
    author = "Aid, S. and others",
    collaboration ="H1 collaboration",
    title = "Elastic photoproduction of $\rho^0$ mesons at \uppercase{HERA}",
    eprint = "hep-ex/9601004",
    archivePrefix = "arXiv",
    reportNumber = "DESY-95-251",
    doi = "10.1016/0550-3213(96)00045-4",
    journal = "Nucl. Phys.",
    volume = "B463",
    pages = "3--32",
    year = "1996"
}

@article{Guzey:2016qwo,
    author = "Guzey, V. and Strikman, M. and Zhalov, M.",
    title = "Accessing transverse nucleon and gluon distributions in heavy nuclei using coherent vector meson photoproduction at high energies in ion ultraperipheral collisions",
    eprint = "1611.05471",
    archivePrefix = "arXiv",
    primaryClass = "hep-ph",
    doi = "10.1103/PhysRevC.95.025204",
    journal = "Phys. Rev.",
    volume = "C95",
    number = "2",
    pages = "025204",
    year = "2017"
}

@article{Baur:2001jj,
    author = "Baur, Gerhard and Hencken, Kai and Trautmann, Dirk and Sadovsky, Serguei and Kharlov, Yuri",
    title = "Coherent $\gamma\gamma$ and $\gamma$-\uppercase{A} interactions in very peripheral collisions at relativistic ion colliders",
    eprint = "hep-ph/0112211",
    archivePrefix = "arXiv",
    doi = "10.1016/S0370-1573(01)00101-6",
    journal = "Phys. Rept.",
    volume = "364",
    pages = "359--450",
    year = "2002"
}

@article{ALICE:2023jgu,
    author = "Acharya, Shreyasi and others",
    collaboration = "ALICE colaboration",
    title = "Energy dependence of coherent photonuclear production of \ensuremath{J/\psi} mesons in ultra-peripheral \uppercase{P}b-\uppercase{P}b collisions at $ \sqrt{{\textrm{s}}_{\textrm{NN}}} $ = 5.02 \uppercase{T}e\uppercase{V}",
    eprint = "2305.19060",
    archivePrefix = "arXiv",
    primaryClass = "nucl-ex",
    reportNumber = "CERN-EP-2023-100",
    doi = "10.1007/JHEP10(2023)119",
    journal = "JHEP",
    volume = "10",
    pages = "119",
    year = "2023"
}

@article{Frankfurt:2015cwa,
    author = "Frankfurt, L. and Guzey, V. and Strikman, M. and Zhalov, M.",
    title = "Nuclear shadowing in photoproduction of \ensuremath{\rho} mesons in ultraperipheral nucleus collisions at \uppercase{RHIC} and the \uppercase{LHC}",
    eprint = "1506.07150",
    archivePrefix = "arXiv",
    primaryClass = "hep-ph",
    doi = "10.1016/j.physletb.2015.11.012",
    journal = "Phys. Lett.",
    volume = "B752",
    pages = "51--58",
    year = "2016"
}

@article{Mantysaari:2023xcu,
    author = {M\"antysaari, Heikki and Salazar, Farid and Schenke, Bj\"orn},
    title = "Energy dependent nuclear suppression from gluon saturation in exclusive vector meson production",
    eprint = "2312.04194",
    archivePrefix = "arXiv",
    primaryClass = "hep-ph",
    reportNumber = "INT-PUB-24-008",
    doi = "10.1103/PhysRevD.109.L071504",
    journal = "Phys. Rev.",
    volume = "D109",
    number = "7",
    pages = "L071504",
    year = "2024"
}

@article{ZEUS:1996zse,
    author = "Derrick, M. and others",
    collaboration ="ZEUS collaboration",
    title = "Measurement of elastic $\omega$ photoproduction at \uppercase{HERA}",
    eprint = "hep-ex/9608010",
    archivePrefix = "arXiv",
    reportNumber = "DESY-96-159",
    doi = "10.1007/s002880050297",
    journal = "Z. Phys.",
    volume = "C73",
    pages = "73--84",
    year = "1996"
}

@article{Gribov:1968jf,
    author = "Gribov, V. N.",
    title = "Glauber corrections and the interaction between high-energy hadrons and nuclei",
    journal = "Sov. Phys. JETP",
    volume = "29",
    pages = "483--487",
    year = "1969"
}

@article{Guzey:2013jaa,
    author = "Guzey, V. and Strikman, M. and Zhalov, M.",
    title = "Disentangling coherent and incoherent quasielastic \ensuremath{J/\psi} photoproduction on nuclei by neutron tagging in ultraperipheral ion collisions at the \uppercase{LHC}",
    eprint = "1312.6486",
    archivePrefix = "arXiv",
    primaryClass = "hep-ph",
    doi = "10.1140/epjc/s10052-014-2942-z",
    journal = "Eur. Phys. J.",
    volume = "C74",
    number = "7",
    pages = "2942",
    year = "2014"
}

@article{ALICE:2021tyx,
    author = "Acharya, Shreyasi and others",
    collaboration = "ALICE colaboration",
    title = "First measurement of the $|t|$-dependence of coherent \ensuremath{J/\psi} photonuclear production",
    eprint = "2101.04623",
    archivePrefix = "arXiv",
    primaryClass = "nucl-ex",
    reportNumber = "CERN-EP-2021-003",
    doi = "10.1016/j.physletb.2021.136280",
    journal = "Phys. Lett.",
    volume = "B817",
    pages = "136280",
    year = "2021"
}

@article{ALICE:2015nbw,
    author = "Adam, Jaroslav and others",
    collaboration = "ALICE colaboration",
    title = "Coherent $\rho^{0}$ photoproduction in ultra-peripheral \uppercase{P}b-\uppercase{P}b collisions at $\sqsnn=2.76 $ \uppercase{T}e\uppercase{V}",
    eprint = "1503.09177",
    archivePrefix = "arXiv",
    primaryClass = "nucl-ex",
    reportNumber = "CERN-PH-EP-2015-082",
    doi = "10.1007/JHEP09(2015)095",
    journal = "JHEP",
    volume = "09",
    pages = "095",
    year = "2015"
}

@article{Soding:1965nh,
    author = "Soding, P.",
    title = "On the apparent shift of the rho meson mass in photoproduction",
    doi = "10.1016/0031-9163(66)90451-3",
    journal = "Phys. Lett.",
    volume = "19",
    pages = "702--704",
    year = "1966"
}

@article{ALICE:2018oyo,
    author = "Acharya, Shreyasi and others",
    collaboration = "ALICE colaboration",
    title = "Energy dependence of exclusive $\mathrm {J}/\psi $ photoproduction off protons in ultra-peripheral p\textendash{}\uppercase{P}b collisions at $\sqrt{s_{\mathrm {\scriptscriptstyle NN}}} = 5.02$ \uppercase{T}e\uppercase{V}",
    eprint = "1809.03235",
    archivePrefix = "arXiv",
    primaryClass = "nucl-ex",
    reportNumber = "CERN-EP-2018-236",
    doi = "10.1140/epjc/s10052-019-6816-2",
    journal = "Eur. Phys. J.",
    volume = "C79",
    number = "5",
    pages = "402",
    year = "2019"
}

@article{ALICE:2021gpt,
    author = "Acharya, Shreyasi and others",
    collaboration = "ALICE colaboration",
    title = "Coherent \ensuremath{J/\psi} and \ensuremath{\psi'} photoproduction at midrapidity in ultra-peripheral \uppercase{P}b-\uppercase{P}b collisions at $\sqsnn = 5.02$ \uppercase{T}e\uppercase{V}",
    eprint = "2101.04577",
    archivePrefix = "arXiv",
    primaryClass = "nucl-ex",
    reportNumber = "CERN-EP-2021-002",
    doi = "10.1140/epjc/s10052-021-09437-6",
    journal = "Eur. Phys. J.",
    volume = "C81",
    number = "8",
    pages = "712",
    year = "2021"
}

@article{CMS:2023snh,
    author = "Tumasyan, Armen and others",
    collaboration ="CMS collaboration",
    title = "Probing small \uppercase{B}jorken-x nuclear gluonic structure via coherent \ensuremath{J/\psi} photoproduction in ultraperipheral \uppercase{P}b-\uppercase{P}b collisions at 
$\sqrt{s_{\mathrm {\scriptscriptstyle NN}}} = 5.02$
\,\uppercase{T}e\uppercase{V}",
    eprint = "2303.16984",
    archivePrefix = "arXiv",
    primaryClass = "nucl-ex",
    reportNumber = "CMS-HIN-22-002, CERN-EP-2023-031",
    doi = "10.1103/PhysRevLett.131.262301",
    journal = "Phys. Rev. Lett.",
    volume = "131",
    number = "26",
    pages = "262301",
    year = "2023"
}

@article{STAR:2023vvb,
    author = "Abdulhamid, M. I. and others",
    collaboration ="STAR collaboration",
    title = "Exclusive \ensuremath{J/\psi}, \ensuremath{\psi(2S)}, and $e^+e^\ensuremath{-}$ pair production in \uppercase{A}u+\uppercase{A}u ultraperipheral collisions at the \uppercase{BNL} \uppercase{R}elativistic \uppercase{H}eavy \uppercase{I}on \uppercase{C}ollider",
    eprint = "2311.13632",
    archivePrefix = "arXiv",
    primaryClass = "nucl-ex",
    doi = "10.1103/PhysRevC.110.014911",
    journal = "Phys. Rev.",
    volume = "C110",
    number = "1",
    pages = "014911",
    year = "2024"
}

@article{Davier:2017zfy,
    author = "Davier, Michel and Hoecker, Andreas and Malaescu, Bogdan and Zhang, Zhiqing",
    title = "Reevaluation of the hadronic vacuum polarisation contributions to the \uppercase{S}tandard \uppercase{M}odel predictions of the muon $g-2$ and ${\alpha (m_Z^2)}$ using newest hadronic cross-section data",
    eprint = "1706.09436",
    archivePrefix = "arXiv",
    primaryClass = "hep-ph",
    doi = "10.1140/epjc/s10052-017-5161-6",
    journal = "Eur. Phys. J.",
    volume = "C77",
    number = "12",
    pages = "827",
    year = "2017"
}

@article{LHCb:2008vvz,
    author = "Alves, Jr., A. Augusto and others",
    collaboration ="LHCb collaboration",
    title = "The \uppercase{LHC}b detector at the \uppercase{LHC}",
    reportNumber = "LHCb-DP-2008-001",
    doi = "10.1088/1748-0221/3/08/S08005",
    journal = "JINST",
    volume = "3",
    pages = "S08005",
    year = "2008"
}

@article{LHCb:2014set,
    author = "Aaij, Roel and others",
    collaboration ="LHCb collaboration",
    title = "\uppercase{LHC}b detector performance",
    eprint = "1412.6352",
    archivePrefix = "arXiv",
    primaryClass = "hep-ex",
    reportNumber = "LHCB-DP-2014-002, CERN-PH-EP-2014-290",
    doi = "10.1142/S0217751X15300227",
    journal = "Int. J. Mod. Phys.",
    volume = "A30",
    number = "07",
    pages = "1530022",
    year = "2015"
}

@article{Spital:1974cx,
    author = "Spital, R. and Yennie, D. R.",
    title = "$\rho^0$ shape in photoproduction",
    doi = "10.1103/PhysRevD.9.126",
    journal = "Phys. Rev.",
    volume = "D9",
    pages = "126--137",
    year = "1974"
}

@article{Jegerlehner:2011ti,
    author = "Jegerlehner, Fred and Szafron, Robert",
    title = "$\rho^0 - \gamma$ mixing in the neutral channel pion form factor $F_{\pi}^{e}(s)$ and its role in comparing $e^+ e^-$ with $\tau$ spectral functions",
    eprint = "1101.2872",
    archivePrefix = "arXiv",
    primaryClass = "hep-ph",
    reportNumber = "DESY-11-008, HU-EP-11-04",
    doi = "10.1140/epjc/s10052-011-1632-3",
    journal = "Eur. Phys. J.",
    volume = "C71",
    pages = "1632",
    year = "2011"
}

@article{Bartos:2017ils,
    author = "Barto\v{s}, Erik and Dubni\v{c}ka, Stanislav and Liptaj, Andrej and Dubni\v{c}kov\'a, Anna Zuzana and Kami\'nski, Robert",
    title = "What are the correct $\rho^0$(770) meson mass and width values?",
    doi = "10.1103/PhysRevD.96.113004",
    journal = "Phys. Rev.",
    volume = "D96",
    number = "11",
    pages = "113004",
    year = "2017"
}

@article{Akiba:2018neu,
    author = "Akiba, K. Carvalho and others",
    title = "The \uppercase{H}e\uppercase{RSC}he\uppercase{L} detector: high-rapidity shower counters for \uppercase{LHC}b",
    eprint = "1801.04281",
    archivePrefix = "arXiv",
    primaryClass = "physics.ins-det",
    reportNumber = "LHCB-DP-2016-003",
    doi = "10.1088/1748-0221/13/04/P04017",
    journal = "JINST",
    volume = "13",
    number = "04",
    pages = "P04017",
    year = "2018"
}

@article{LHCb:2014vhh,
    author = "Aaij, Roel and others",
    collaboration ="LHCb collaboration",
    title = "Precision luminosity measurements at \uppercase{LHC}b",
    eprint = "1410.0149",
    archivePrefix = "arXiv",
    primaryClass = "hep-ex",
    reportNumber = "LHCB-PAPER-2014-047, CERN-PH-EP-2014-221",
    doi = "10.1088/1748-0221/9/12/P12005",
    journal = "JINST",
    volume = "9",
    number = "12",
    pages = "P12005",
    year = "2014"
}

@article{GEANT4:2002zbu,
    author = "Agostinelli, S. and others",
    collaboration = "Geant4 collaboration",
    title = "GEANT4--a simulation toolkit",
    reportNumber = "SLAC-PUB-9350, FERMILAB-PUB-03-339, CERN-IT-2002-003",
    doi = "10.1016/S0168-9002(03)01368-8",
    journal = "Nucl. Instrum. Meth.",
    volume = "A506",
    pages = "250--303",
    year = "2003"
}

@article{Clemencic:2011zza,
    author = "Clemencic, M. and Corti, G. and Easo, S. and Jones, C. R. and Miglioranzi, S. and Pappagallo, M. and Robbe, P.",
    editor = "Lin, Simon C.",
    collaboration ="LHCb collaboration",
    title = "The \uppercase{LHC}b simulation application, \uppercase{G}auss: Design, evolution and experience",
    doi = "10.1088/1742-6596/331/3/032023",
    journal = "J. Phys. Conf. Ser.",
    volume = "331",
    pages = "032023",
    year = "2011"
}

@article{Antcheva:2011zz,
    author = "Antcheva, I. and others",
    title = "ROOT: A \uppercase{C}++ framework for petabyte data storage, statistical analysis and visualization",
    reportNumber = "FERMILAB-PUB-11-930-SCD",
    doi = "10.1016/j.cpc.2011.02.008",
    journal = "Comput. Phys. Commun.",
    volume = "182",
    pages = "1384--1385",
    year = "2011"
}

@article{CMS:2019awk,
    author = "Sirunyan, Albert M and others",
    collaboration ="CMS collaboration",
    title = "Measurement of exclusive $\rho(770)^0$ photoproduction in ultraperipheral p\uppercase{P}b collisions at $\sqrt{s_\mathrm{NN}} =$ 5.02 \uppercase{T}e\uppercase{V}",
    eprint = "1902.01339",
    archivePrefix = "arXiv",
    primaryClass = "hep-ex",
    reportNumber = "CMS-FSQ-16-007, CERN-EP-2018-285",
    doi = "10.1140/epjc/s10052-019-7202-9",
    journal = "Eur. Phys. J.",
    volume = "C79",
    number = "8",
    pages = "702",
    year = "2019"
}

@article{Brodsky:1994kf,
    author = "Brodsky, Stanley J. and Frankfurt, L. and Gunion, J. F. and Mueller, Alfred H. and Strikman, M.",
    title = "Diffractive leptoproduction of vector mesons in \uppercase{QCD}",
    eprint = "hep-ph/9402283",
    archivePrefix = "arXiv",
    reportNumber = "SLAC-PUB-6412, CU-TP-617, UCD-93-36",
    doi = "10.1103/PhysRevD.50.3134",
    journal = "Phys. Rev.",
    volume = "D50",
    pages = "3134--3144",
    year = "1994"
}

@article{Klein:2019qfb,
    author = {Klein, Spencer R. and M\"antysaari, Heikki},
    title = "Imaging the nucleus with high-energy photons",
    eprint = "1910.10858",
    archivePrefix = "arXiv",
    primaryClass = "hep-ex",
    doi = "10.1038/s42254-019-0107-6",
    journal = "Nature Rev. Phys.",
    volume = "1",
    number = "11",
    pages = "662--674",
    year = "2019"
}

@article{Guzey:2020ntc,
    author = "Guzey, V. and Kryshen, E. and Strikman, M. and Zhalov, M.",
    title = "Nuclear suppression from coherent \ensuremath{J/\psi} photoproduction at the Large Hadron Collider",
    eprint = "2008.10891",
    archivePrefix = "arXiv",
    primaryClass = "hep-ph",
    doi = "10.1016/j.physletb.2021.136202",
    journal = "Phys. Lett.",
    volume = "B816",
    pages = "136202",
    year = "2021"
}

@article{Bertulani:2005ru,
    author = "Bertulani, Carlos A. and Klein, Spencer R. and Nystrand, Joakim",
    title = "Physics of ultra-peripheral nuclear collisions",
    eprint = "nucl-ex/0502005",
    archivePrefix = "arXiv",
    doi = "10.1146/annurev.nucl.55.090704.151526",
    journal = "Ann. Rev. Nucl. Part. Sci.",
    volume = "55",
    pages = "271--310",
    year = "2005"
}

@article{Jones:2014aoa,
    author = "Jones, Adam B. and Brown, B. Alex",
    title = "Two-parameter \uppercase{F}ermi function fits to experimental charge and point-proton densities for $^{208}$\uppercase{P}b",
    doi = "10.1103/PhysRevC.90.067304",
    journal = "Phys. Rev.",
    volume = "C90",
    number = "6",
    pages = "067304",
    year = "2014"
}

@article{Klein:2016dtn,
    author = "Klein, Spencer R.",
    collaboration ="STAR collaboration",
    title = "Ultra-peripheral collisions with gold ions in \uppercase{STAR}",
    eprint = "1606.02754",
    archivePrefix = "arXiv",
    primaryClass = "nucl-ex",
    doi = "10.22323/1.265.0188",
    journal = "PoS",
    volume = "DIS2016",
    pages = "188",
    year = "2016"
}

@article{Davier:2019can,
    author = "Davier, M. and Hoecker, A. and Malaescu, B. and Zhang, Z.",
    title = "A new evaluation of the hadronic vacuum polarisation contributions to the muon anomalous magnetic moment and to $\alpha(m_Z^2)$",
    eprint = "1908.00921",
    archivePrefix = "arXiv",
    primaryClass = "hep-ph",
    doi = "10.1140/epjc/s10052-020-7792-2",
    journal = "Eur. Phys. J.",
    volume = "C80",
    number = "3",
    pages = "241",
    year = "2020",
    extraPrefix    = "Erratum",
      extraVolume    = "C80",
      extraPages     = "410",
      extraYear      = "2020",
      extraDoi       = "10.1140/epjc/s10052-020-7792-2",
}

@article{Chew:1952fca,
    author = "Chew, Geoffrey F. and Wick, Gian Carlo",
    title = "The impulse approximation",
    doi = "10.1103/PhysRev.85.636",
    journal = "Phys. Rev.",
    volume = "85",
    number = "4",
    pages = "636",
    year = "1952"
}

@article{Tarbert:2013jze,
    author = "Tarbert, C. M. and others",
    title = "{Neutron skin of $^{208}$\uppercase{P}b from coherent pion photoproduction}",
    eprint = "1311.0168",
    archivePrefix = "arXiv",
    primaryClass = "nucl-ex",
    doi = "10.1103/PhysRevLett.112.242502",
    journal = "Phys. Rev. Lett.",
    volume = "112",
    number = "24",
    pages = "242502",
    year = "2014"
}

@article{BaBar:2012bdw,
    author = "Lees, J. P. and others",
    collaboration = "\uppercase{B}a\uppercase{B}ar collaboration",
    title = "Precise measurement of the $e^+ e^- \to \pi^+\pi^- (\gamma)$ cross section with the initial-state radiation method at \uppercase{B}a\uppercase{B}ar",
    eprint = "1205.2228",
    archivePrefix = "arXiv",
    primaryClass = "hep-ex",
    reportNumber = "BABAR-PUB-12-003",
    doi = "10.1103/PhysRevD.86.032013",
    journal = "Phys. Rev.",
    volume = "D86",
    pages = "032013",
    year = "2012"
}

@book{Donnachie:2002en,
    author = "Donnachie, S. and Dosch, Hans Gunter and Nachtmann, O. and Landshoff, P.",
    title = "Pomeron physics and \uppercase{QCD}",
    isbn = "978-0-511-06050-2, 978-0-521-78039-1, 978-0-521-67570-3",
    publisher = "Cambridge University Press",
    volume = "19",
    month = "12",
    year = "2004"
}

@article{Ioffe:1969kf,
    author = "Ioffe, B. L.",
    title = "Space-time picture of photon and neutrino scattering and electroproduction cross-section asymptotics",
    doi = "10.1016/0370-2693(69)90415-8",
    journal = "Phys. Lett.",
    volume = "B30",
    pages = "123--125",
    year = "1969"
}


%% file: standard.bib
@article{PDG2024,
     author    = "Navas, S. and others",
    collaboration = "Particle Data Group",
     title     = "{\href{http://pdg.lbl.gov/}{Review of particle physics}}",
     journal   = "Phys. Rev.",
     year = {2024},
     number = {8},
     volume      = "D110",
     pages     = "030001",
     doi = "10.1103/PhysRevD.110.030001"
}

@article{Allison:2006ve,
      author         = "Allison, John and Amako, K. and Apostolakis, J. and
                        Araujo, H. and Dubois, P.A. and others",
 collaboration = "Geant4 collaboration",
      title          = "{Geant4 developments and applications}",
      journal        = "IEEE Trans.Nucl.Sci.",
      volume         = "53",
      pages          = "270",
      doi            = "10.1109/TNS.2006.869826",
      year           = "2006",
      reportNumber   = "SLAC-PUB-11870",
}

@Misc{mciteplus,
  author = 	 {Shell, Michael},
  title = 	 {Mciteplus: Enhanced multicitations},
  howpublished = {\href{http://www.michaelshell.org/tex/mciteplus/} {http://www.michaelshell.org/tex/mciteplus/}},
}
